\documentclass{article}
\begin{document}

\title{RANDOM WALKS ON GRAPHS: IDEAS, TECHNIQUES AND RESULTS}         
\author{ R. Burioni and D. Cassi\\ \\
      Dipartimento di Fisica, INFM and INFN \\
     Universit\`a di Parma, \\ Parco Area delle Scienze 7/A, 43100 Parma, Italy}
\date{}          
\maketitle


\begin{abstract}
Random walks on graphs are widely used in all sciences to describe a great variety of phenomena where dynamical random processes are affected by topology. In recent years, relevant mathematical results have been obtained in this field, and new ideas have been introduced, which can be fruitfully extended to different areas and disciplines. Here we aim at giving a brief but comprehensive perspective of these progresses, with a particular emphasis on physical aspects.   
\end{abstract}
\tableofcontents
\section{Introduction}

A graph is the most general mathematical description of a set of elements connected pairwise by some kind of relation. Therefore, it is not surprising that graph theory has been successfully applied to a wide range of very different disciplines, from biology to social science, computing, psychology, economy, chemistry and physics.

\noindent
In recent times, physicists have been mainly interested in graphs as models of complex systems, in condensed matter and in network theory. Indeed, these structures have proven to be very useful to describe inhomogeneous structures such as disordered materials, glasses, polymers, biomolecules  as well as electric circuits, communication networks, statistical models of algorithms, and applications of statistical mechanics to different (non physical) systems.

The function of graphs in physics, however, is not purely descriptive. Geometry and topology have a deep influence on the physical properties of complex systems, where the presence of a large number of interacting degrees of freedom typically matters more than the interaction details. In fact, the specific interest of a physicist concerns the properties of a graph which most affect the dynamical and thermodynamical behaviour of the system it describes. 
On the other hand, the study of complex systems requires the introduction of statistical methods, to give an effective description of a number of quantities which, otherwise, would be too difficult to control. 

\noindent
Random walks are probably the simplest stochastic process affected by topology and, at the same time, the basic model of diffusion phenomena and non-deterministic motion.  
They have been extensively studied for decades on regular structures such as lattices, and most of the common wisdom concerning them relies on the results obtained in this particular geometry. The richer topology of a generic graph can have a dramatic effect on the properties of random walks, especially when considering infinite graphs, which are introduced to describe macroscopic systems in the thermodynamic limit. There, the asymptotic behaviour at long time typically exhibits universal features, only depending on large scale topology. On lattices, such features are known to be related to the Euclidean dimension only. On general graphs, universality allows to generalize the concept of dimension to inhomogeneous structures, providing a very powerful tool to investigate a large class of different physical models, apparently not connected to diffusion processes. On the other hand, a new and unexpected phenomenon arises in presence of strong inhomogeneity, namely the splitting between local and average properties. This provides a fundamental conceptual framework to investigate complex systems even from an experimental point of view.

\noindent
Most results concerning random walks on graphs in physics have been obtained in the last two decades and are scattered in a large number of technical papers. This review is intended to provide the reader with a rigorous, self-contained and up to date account of the present knowledge about this subject. Particular attention has been paid to give a simple and general framework effectively resuming rather different results. As for specific calculations, we refer to bibliography, unless required by clarity reasons. The emphasis is always put on the physical meaning. The reader more interested in formal aspects can find a presentation focused on mathematics in another recent review \cite{vienna}.

\noindent
The paper is organized as follows:
In the first sections we give a brief mathematical description of graphs and random walks, introducing the language and the formalism we will use through the whole article. Then we present a simple treatment of the finite graphs case, before dealing with infinite graphs. The latter require the introduction of specific concepts, which are fully discussed in an introductory section. Then, the asymptotic behaviour of random walks on infinite graphs is studied and used to define the type problem and the spectral dimension. The difference between local and average properties is evidenced in the following sections. The concluding chapters are devoted to the analysis of a large class of specific graphs and to the relations of random walks with different physical problems.

\section{Mathematical description of graphs}

Let us begin by  introducing the basic mathematical definitions and results concerning graphs \cite{harary}.

\noindent 
A graph $\cal{G}$ is a countable set $V$ of vertices (or sites)  $(i)$ 
connected pairwise by a set $E$ of unoriented links (or bonds) $(i,j)=(j,i)$.  
If the set $V$ is finite, ${\cal{G}}$ is called a finite graph and we will denote by $N$ 
the number of vertices of $\cal{G}$. Otherwise, when $V$ is infinite, ${\cal{G}}$ is called an infinite graph.
A subgraph $\cal{S}$ of $\cal{G}$ is a graph whose set of vertices $S\subseteq V$ and whose set of links $E' \subseteq E$.

\noindent
A path $C_{i\to j}$ in $\cal {G}$ connecting points $i$ and $j$ is a sequence of consecutive links $\{(i,k)(k,h)\dots(n,m)(m,j)\}$ 
and a graph is said to be connected, if for any two points $i,j \in V$ 
there is always a path joining them. In the following we will consider only connected graphs.
 
\noindent
Every connected graph $\cal{G}$ is endowed with an intrinsic metric 
generated by the chemical distance $r_{ij}$ which is defined as the number of links in the shortest path 
connecting vertices $i$ and $j$. 

\noindent
A particular class of graphs, often occurring in physical applications, is characterized by the absence of closed paths containing an odd number of links. These graphs are called {\it bipartite}, since we can divide their sites into $2$ sets $V_1$ and $V_2$ such that the points of $V_1$ are connected by a link only to points $V_2$ and viceversa. Square and hypercubyc lattices are the most typical examples of bipartite graphs, as well as all trees (graphs without closed self-avoiding paths).

\noindent
The graph topology can be algebraically represented introducing its adjacency matrix $A_{ij}$ given by:
\begin{equation}
A_{ij}=\left\{
\begin{array}{cl}
1 & {\rm if } \ (i,j) \in E \cr
0 & {\rm if } \ (i,j) \not\in E \cr
\end{array}
\right .
\label{defA}
\end{equation}

\noindent
The Laplacian matrix $\Delta_{ij}$ is defined by:
\begin{equation}
\Delta_{ij} = z_i \delta_{ij} - A_{ij}
\label{defDelta}
\end{equation}
where $z_i=\sum_j A_{ij}$, the number of nearest neighbours of $i$, 
is called the coordination number of site $i$. 

\noindent
In order to describe disordered structures we introduce 
a generalization of the adjacency matrix  given by the ferromagnetic
coupling matrix $J_{ij}$, with $J_{ij} \neq 0 \iff A_{ij}=1$ and
$\sup_{(i,j)} J_{ij} < \infty$, $\inf_{(i,j)} J_{ij} > 0$. 
One can then define the generalized 
Laplacian:
\begin{equation}
L_{ij} = I_i \delta_{ij} - J_{ij}
\label{defL}
\end{equation}
where $ I_i =\sum_j J_{ij}$.

\section{The random walk problem}       

Let us now introduce the so called {\it simple random walk} on a graph $\cal{G}$.
Assuming the time $(t)$ to be discrete, we define at each time step $t$  the jumping 
probability $p_{ij}$ between nearest neighbour sites $i$ and $j$:
\begin{equation}
p_{ij}={A_{ij}\over z_i} =(Z^{-1}A)_{ij}
\label{pij}
\end{equation}
where $Z_{ij}= z_i \delta_{ij}$. 

\noindent
This is the simplest case we can consider: the jumping probabilities are isotropic at each point and they do not depend on time; in addition the walker is forced to jump at every time step. As we will see later, the last condition, i.e. the impossibility of staying on site, although crucial for the short time behaviour, has no significant influence on the long time regime.

\noindent
Usually, the  random walk problem is considered to be completely solvable if, for any $i,j\in \cal{G}$ and $t\in N$, we are able to calculate the functions $P_{ij}(t)$, each representing the probability of being in site $j$ at time $t$ for a walker starting from site $i$ at time $0$. These probabilities are the elements of a matrix $P=||P_{ij}(t)||$ which is equal to the t-th power of the jumping probabilities matrix $p=||p_{ij}||$:
\begin{equation}
P_{ij}(t)=(p^t)_{ij}~~.
\label{ptij}
\end{equation}
The relation (\ref{ptij}) can be easily proven by induction on $t$. It also has an interesting physical interpretation as a sum over paths; developing the matrix products term by term we can write the whole expression as
\begin{equation}
P_{ij}(t)=(p^t)_{ij}=\sum_{C_{i\to j}(t)}w(C_{i\to j}(t))
\label{sommacammini}
\end{equation}
where the sum is over all $t-$steps paths between $i$ and $j$. 
The weight $w(C_{i\to j}$ is the probability for the walker of going from $i$ to $j$ following exactly the path $w(C_{i\to j})$:
\begin{equation}
w(C_{i\to j}(t))=\prod_{(k,l)\in C_{i\to j}(t)}p_{kl}
\label{pesocammino}
\end{equation}
the product being over all the $t$ links belonging to the path.

\noindent
The calculation of all $P_{ij}(t)$, which is straightforward as far as relatively small graphs are concerned, for large or infinite graphs becomes practically impossible and, above all, little significant. In fact, for large systems we are mainly interested in global and collective properties as it typically happens in statistical physics. Therefore, a small subset of all these quantities is usually chosen, together with some other related to them, which give an effective physical description of the random walker behaviour.
The most relevant of them is from many points of view the probability $P_{ii}(t)$ of returning to the starting point after $t$ steps, also called the {\it random walk autocorrelation function}. As we will see, its asymptotic behaviour gives the most direct characterization of the large scale topology for infinite graphs. 
A related quantity is the average number  $P_{ii}$ of returns to the starting point $i$, which can be generalized to the average number $P_{ij}$ of passages through $j$ starting from $i$:
\begin{equation}
P_{ij}\equiv\lim_{t\to\infty} \sum_{k=0}^t P_{ij}(k),
\end{equation}
where the limit can be infinite.

\noindent
The mean displacement $r_i(t)$ from the starting site $i$ after $t$ steps is deeply related to the diffusion properties and is defined as
\begin{equation}
r_i(t)\equiv \sum_j r_{ij} P_{ij}(t)
\label{spostamentomedio}
\end{equation}
Notice that, unlike the case of random walks in continuous Euclidean space, here we consider $r$ instead of $r^2$, the latter having no particular significance in absence of Euclidean metric. 

\noindent
The quantities introduced up to now are not "sensible to the  history". Indeed, we can in principle determine all of them simply by considering the situation of the walker at time $t$ regardless of his previous behaviour. In order to keep track of what happened before the instant $t$, a different class of functions is introduced, starting with the {\it first passage probability} 
$F_{ij}(t)$. The latter denotes the conditional probability for a walker starting from $i$ of reaching
for the first time the site $j\not = i$ in $t$ steps. For $i=j$ the previous definition would not be interesting, being the walker in $i$ at $t=0$ by definition.  Therefore, one defines  $F_{ii}(t)$ 
to be the probability of returning to the starting point $i$ for the first time 
after $t$ steps and one sets $F_{ii}(0)=0$.  
In spite of the deeply different nature of $P$ and $F$, a fundamental relation can be established between them if all time steps form $0$ to $t$ are taken into account (in other words, we have to give up the time locality): 
\begin{equation}
P_{ij}(t)=\sum_{k=0}^t F_{ij}(k) P_{jj}(t-k) + \delta_{ij}\delta_{t 0}.
\label{PF}
\end{equation}
This can be easily obtained by considering that each walker which is in $j$ at time $t$ only has two possibilities: either it gets there for the first time, or it has reached $j$ for the first time at a previous time $k$ and then it has returned there after $t-k$ steps.
The first passage probability is in turn connected to other meaningful history dependent quantities.
The probability $F_{ij}$ of ever reaching the site $j$ starting from $i$ (or of ever returning to i, if $i=j$) is given by
\begin{equation}
F_{ij}=\sum_{t=0}^\infty F_{ij}(t)
\label{Fij}
\end{equation} 

\noindent
By $S_i(t)$ we denote the average number of different sites visited after $t$ steps by a walker starting from $i$. Its relation to $F_{ij}(t)$ is 
\begin{equation}
S_{i}(t)=1+\sum_{k=1}^t \sum_j F_{ij}(k)
\label{Sit}
\end{equation} 
Finally, the {\it first passage time} $t_{ij}$, i.e. the average time at which a walker starting from $i$, and passing at least once through $j$, reaches $j$ for the first time
(or returns for the first time to $i$, if $i=j$) is
\begin{equation}
t_{ij}=\lim_{t\to\infty}{\sum_{k=0}^t k F_{ij}(k) \over F_{ij}}
\label{tij}
\end{equation}

\noindent
The simple random walk  can be modified to give a richer behaviour and to describe more general physical problems. Indeed, one can  
introduce anisotropic jumping probabilities by substituting in (\ref{pij}) the adjacency matrix with a ferromagnetic coupling matrix:
\begin{equation}
p_{ij}={J_{ij}\over I_i} =(I^{-1}J)_{ij}
\label{pijferr}
\end{equation}
where $I_{ij}= I_i \delta_{ij}$ and $I_i=\sum_{k}J_{ik}$. Depending on the specific properties of $J_{ij}$, this can produce only local effects or introduce a global bias which destroys the leading diffusive behaviour  giving rise to transport phenomena. 

\noindent
Moreover, one can relax the constraint of jumping at each time step by introducing waiting and traps on the sites.
The jumping  probabilities are then modified to:
\begin{equation}
  p_{i,j} = \frac{J_{i,j}+w_i\delta_{i,j}}{I_i+w_i+d_i} 
\label{wt}
\end{equation}
where both $w_i$ and $d_i$ are real positive numbers. From (\ref{wt}), $w_i/(I_i+w_i+d_i)$ is the probability for the walker to stay on site $i$ instead of jumping away and  $d_i/(I_i+w_i+d_i)$ the probability of disappearing (or dying, or being trapped forever) at site $i$.
As we will see later, waiting only affects the short time behaviour, while traps can dramatically modify also the long time asymptotic properties.

\section{The generating functions} 
      
Even if we consider only the few fundamental quantities devised at the end of the last section, their direct calculation can be in practice a hard or impossible task on general graphs. However, a powerful indirect mathematical technique exists allowing overcoming a series of typical difficulties: this is the discrete Laplace transform, which maps a time function into its {\it generating function}.    
The generating function $\tilde{f}(\lambda)$ of $f(t)$is defined by:
\begin{equation}
\tilde{f}(\lambda)=\sum_{t=0}^{\infty}\lambda^t f(t)
\label{genfun}
\end{equation}
where $\lambda $ is a complex number.
The inverse equation giving $f(t)$ from $\tilde{f}(\lambda)$ is 
\begin{equation}
f(t)={\partial^t \tilde{f}(\lambda) \over \partial \lambda^t}\bigg\vert_{\lambda=0}.
\label{invgenfun}
\end{equation}
This equation is useful as fare as we are interested in small $t$ behaviour, but it becomes absolutely ineffectual in the study of asymptotic regimes for $t\to\infty$. In this case a very powerful tool is provided by the Tauberian theorems, relating the singularities of $\tilde{f}(\lambda)$ to the leading large $t$ behaviour of $f(t)$.
We give here a rather general Tauberian theorem, which is particularly useful when dealing with random walks. The main assumption we make concerns the analytical form of the leading singularity: we only consider power laws and logarithmic behaviours, since all cases discussed in this paper as well as all physically meaningful cases belong to this class.
Suppose that $\tilde{f}(\lambda)$ has its singularity nearest to $\lambda=0$ in $\lambda=1$. 
and that $\tilde f(1-\epsilon)$, for 
$\epsilon \rightarrow 0^+$ goes as
\begin{equation}
\tilde f(1-\epsilon)\sim h(\epsilon)+const \prod _{i=0}^{\infty}\left({}^i
\ln (1/\epsilon)\right)^{\alpha(i)} \label{sette}
\label{toni}
\end{equation}
where ${}^i\ln x\equiv \ln {}^{i-1}\ln x$, with ${}^0\ln x
\equiv x$ e $h(\epsilon)$ is finite for $\epsilon \rightarrow
0^+$.\\
Then, for $t \rightarrow \infty$
\begin{equation}
f(t) \sim  const' r^{-t}\prod_{i=0} ^{\infty} {}^i\ln^{\beta(i)} (t)
\end{equation}
where $\beta(i)$ are related to  $\alpha(i)$ by
\begin{equation}
\beta(i)= \cases{
\alpha(0)-1 & for $i=0$ \cr
\theta (i-m) (\alpha(i)+1) -1-\delta_{i,m}I(\widetilde{d}/2)  & otherwise
}
\end{equation}
where
\begin{equation}
m= {\rm min}
\{i\ge 0 | \beta(i) \neq -1\}
\end{equation}
and
\begin{equation}
I\left( \widetilde {d} / 2 \right) = \cases { 1 & if 
$\widetilde {d} / 2 $ is integer \cr
 0 & otherwise \cr }
\end{equation}

\noindent
The constant $const'$ is in general a function of $const$ and of all the exponents appearing in the previous formulas. We don't give here its rather involved explicit expression, since it is not relevant to the purposes of this paper.

\noindent
The generating functions are usually easier to calculate, since they allow exploiting some peculiar properties of random walks functions. Moreover, a series of relevant random walk parameters which are non-local in time, can be obtained directly from generating function, without calculating the corresponding time dependent quantities. 
A good example is given by $P_{ij}$, $F_{ij}$, and $t_{ij}$ which are related to $\tilde{{P}_{ij}}(\lambda)$ and $\tilde{{F}_{ij}}(\lambda)$ by
\begin{equation}
 P_{ij} = \lim_{\lambda\to1^-}\tilde{P}_{ij}(\lambda)                                          
\label{PP}
\end{equation} 
\begin{equation}
 F_{ij} = \tilde{F}_{ij}(1)                                          
\label{FF}
\end{equation} 
\begin{equation}
 t_{ij} = \lim_{\lambda\to1^-}{\partial \log \tilde{{F}_{ij}}((\lambda) \over \partial \lambda}
\label{tF}
\end{equation}

\noindent
The basic property of random walks generating functions arise from the deconvolution of eq.(\ref{PF}), which after some straightforward steps becomes
\begin{equation}
\tilde{P}_{ij}(\lambda)\tilde{F}_{ij}(\lambda)\tilde{P}_{jj}(\lambda)+\delta_{ij}
\label{gf2}
\end{equation}
In other words, the relation which was non-local in time becomes local in $\lambda$. As we will see in practical application, many  iteration techniques for the analytical calculation of generating functions are based on this property.

\section{Random walks on finite graphs}

Finite graphs consist of a finite number of sites and links. In principle, every physical structure is composed of a finite number of elements, but it is well known that a series of  behaviours occurring in macroscopic systems are better described in the thermodynamic limit.  Indeed, the typical singularities and power laws characterizing phase transitions and asymptotic regimes, such as  large scale, long times, low temperature and low frequency behaviours, can only be found on infinite graphs.

\noindent 
However, finite graphs are appropriate when dealing with mesoscopic structures and finite size effects. 
The random walk problem on finite graphs is simplified by the finiteness of the adjacency matrix. In fact, the analytical study is reducible to a spectral problem on a real finite-dimensional vector space, and numerical simulations are easily implemented by Monte Carlo techniques.

\noindent
Let us first consider the case of random walk without traps, whose jumping probabilities are given by (\ref{wt}) with $d_i=0$ $\forall\quad i$.
The matrix elements $p_{ij}$ satisfy the relations
\begin{equation}
{p}_{ij}\ge 0 \quad\quad\quad \forall\quad i,j
\label{stoc1}
\end{equation}
\begin{equation}
\sum_{j=1}^N {p}_{ij}=1 \quad\quad\quad \forall\quad i        
\label{stoc2}
\end{equation}
defining a {\it stochastic matrix}.

\noindent
The stochastic matrices we are considering exhibits different properties according to the some general features of the graph and  of the jumping probabilities \cite{gantmacher}. We distinguish two cases:
\begin{enumerate}
\item{c1} If $\cal{G}$ is not bipartite, or if it has a staying probability on at least one site, then it has only one eigenvalue $p_{max}$ with maximal modulus and $p_{max}=1$. Moreover, the eigenvector corresponding to $p_{max}$ has the same entry on each site (usually one chooses $v_{max}=(1,1,1,\ldots,1)$ for simplicity)
\item{c2} If $\cal{G}$ is bipartite without staying probabilities, then the spectrum of $p$ is symmetric with respect to the origin of the complex plane. Therefore, in addition to  $p_{max}$ it has a second maximal modulus eigenvalue $p_{min}=-1$. The eigenvector $v_{max}$ has the same properties of the previous case, while $v_{min}$ has all entries on $V_1$ equal to the same number $v$ and all entries on $V_2$ equal to $-v$ (usually one chooses $v=1$)
\end{enumerate}

\noindent
In case c1, one can easily show that the random walk is {\it ergodic}, i.e. that it admits limit probabilities for $t\to\infty$:
\begin{equation}
P_{ij}^\infty=\lim_{t\to\infty}P_{ij}(t)\quad\quad\quad\quad \forall\quad i,j
\label{p limite}
\end{equation}
and that
\begin{equation}
P_{ij}^\infty={1\over N}\quad\quad\quad\quad \forall\quad i,j.
\label{p limite 1}
\end{equation}
This means that, independently of the initial conditions, the asymptotic probabilities are the same over all the graph sites.
Moreover, this uniform limit value is reached exponentially and the exponential decay of each matrix element is no slower than $p_2^t$, $p_2$ being the second greatest eigenvalue of $p_{ij}$.

\noindent
In case c2, the random walk is not ergodic. In particular, we have
\begin{equation}
P_{ij}(t)=0 
\label{p bipartiti}
\end{equation}
for all $t$ such that $t-r_{ij}$ is odd.
On the other hand, considering for each couple of sites $i$ and $j$ only the values $t_{ij}'$ of $t$ having the same parity as $r_{ij}$, 
one can show that 
\begin{equation}
P_{ij}^\infty=\lim_{t_{ij}'\to\infty}P_{ij}(t_{ij}')\quad\quad\quad\quad \forall\quad i,j
\label{p limite bipartiti}
\end{equation}
with
\begin{equation}
P_{ij}^\infty={2\over N}\quad\quad\quad\quad \forall\quad i,j
\label{p limite bipartiti 1}
\end{equation}
and the limit is reached exponentially as in case 1.

\noindent 
Similarly, one can easily prove that 
\begin{equation}
\lim_{t\to\infty}F_{ij}(t)=0 \quad\quad\quad\quad\forall\quad i,j
\end{equation}
and
\begin{equation}
\lim_{t\to\infty}S_{i}(t)=N \quad\quad\quad\quad\forall\quad i
\end{equation}
the limit values being reached exponentially.

\noindent
Moreover,
\begin{equation}
P_{ij}=\infty\quad\quad\quad\quad\forall\quad i,j
\end{equation}
\begin{equation}
F_{ij}=1\quad\quad\quad\quad\forall\quad i,j
\end{equation}
and
\begin{equation}
t_{ij}<\infty \quad\quad\quad\quad\forall\quad i,j
\end{equation}.

\noindent
The introduction of at least one trap, setting $d_i>0$ for at least one site $i$ in (\ref{wt}), dramatically changes the random walk behaviour.
The jumping probabilities matrix $p$ is no longer stochastic, since condition (\ref{stoc2}) is not satisfied.
However, condition (\ref{stoc1}) (i.e. non-negativity) still holds,  implying relevant properties.
We can still distinguish between case $1$ and $2$, but the corresponding properties are modified as follows:
\begin{enumerate}
\item If $\cal{G}$ is not bipartite, or if it has a staying probability on at least one site, then it has only one eigenvalue $p_{max}$ with maximal modulus and $p_{max}<1$. Moreover, the entries $v_{max\, i}$ of the eigenvector $v_{max}$, corresponding to $p_{max}$, have the same sign, and $v_{max}$ is the only eigenvector having such a property.

\item If $\cal{G}$ is bipartite without staying probabilities, then the spectrum of $p$ is symmetric with respect to the origin of the complex plane. Therefore, in addition to  $p_{max}<1$ it has a second maximal modulus eigenvalue $p_{min}=-p_{max}$. The eigenvector $v_{max}$ has the same properties of the previous case, while $v_{min}$ can be chosen in such a way that  all its entries $v_{min\, i}$ on $V_1$ are equal to $v_{max \, i}$  and all its entries $v_{min\, j}$ on $V_2$ equal to $-v_{max \, j}$.
\end{enumerate}

\noindent
In both cases the random walk is ergodic and the limit probabilities vanish:
\begin{equation}
P_{ij}^\infty=0\quad\quad\quad\quad \forall\quad i,j.
\end{equation}
However, in case $2$ the time parity still has to be taken into account and (\ref{p bipartiti}) holds. Moreover, the asymptotic decay is  exponential and no slower than $p_{max}^t$.
Finally, as for the other random walk functions we get
\begin{equation}
\lim_{t\to\infty}F_{ij}(t)=0 \quad\quad\quad\quad\forall\quad i,j
\end{equation}
\begin{equation}
\lim_{t\to\infty}S_{i}(t)<N \quad\quad\quad\quad\forall\quad i
\end{equation}
the limit values being reached exponentially,

\begin{equation}
P_{ij}<\infty\quad\quad\quad\quad\forall\quad i,j
\end{equation}

\begin{equation}
F_{ij}<1\quad\quad\quad\quad\forall\quad i,j
\end{equation}

\begin{equation}
t_{ij}<\infty \quad\quad\quad\quad\forall\quad i,j
\end{equation}.

\section{Infinite graphs}

When dealing with macroscopic systems, composed of a very large number $N$ of sites, one usually takes the thermodynamic limit $N\to \infty$. This means that we have to consider infinite graphs, i.e. graphs composed by an infinite number of sites.
This is particularly convenient for two main reasons:
\begin{itemize}
\item first of all, a single infinite structure effectively describes a very large (infinite, indeed) number of large structures having different sizes, but similar geometrical features
\item the singularities in thermodynamic potentials typical of critical phenomena as well as a series of universal asymptotic behaviours only occur on infinite structures.
\end{itemize}

\noindent
As for random walks on large real structures, the time dependence of physical quantities exhibits different features according to the time scale which is considered.
For very long times, the walker can explore every site and its behaviour is described by the finite graphs laws introduced in the previous section. However, if the time  is long enough to explore large portions of the system, but still too short to experience the finite size effects, many significant quantities are quite insensitive to local details and exhibit power law time dependence with universal exponents. Often, this is the most interesting regime in physical applications. On infinite graphs, this is the true asymptotic regime even for very large times; therefore, we can reproduce the universal behaviours of a huge variety of finite large structures simply by considering infinite graphs with similar topological features.

\noindent
To deal with infinite graphs, some further mathematics has to be introduced. In particular, we need tools to "explore" large scale topology. To this purpose, we define the Generalized Van Hove Spheres (GVHS): 
a GVHS ${\cal{S}}_{o,r}$ of centre $o$ and radius $r$ is the subgraph of $\cal{G}$, given by the set of vertices
$V_{o,r}=\{ i \in V| r_{i,o}\le r \}$ and by the set of links
$E_{o,r}=\{(i,j) \in E| i\in V_{o,r}, j\in V_{o,r}\}$.

\noindent
Let us use  $|S|$ to denote the number of elements of a set
$S$. Then $|V_{o,r}|$, as a function of the distance $r$, describes the growth
rate of the graph at the large scales \cite{woess2}. 
In particular, 
a graph is said to have a {\it polynomial growth} if $\forall o \in V_X \ \exists c,k$, such that
\begin{equation}
|V_{o,r}|<c\ r^k.
\label{growth}
\end{equation}

\noindent
For a graph satisfying (\ref{growth}), we define the upper growth exponent
$d_g^+$ and the lower growth exponent $d^-_g$ as
\begin{equation}
d^+_g=\inf\{k|\ |V_{o,r}|<c_1\ r^k,\forall o \in V\}
\end{equation}
and 
\begin{equation}
d^-_g=\sup\{k|\ |V_{o,r}|>c_2\ r^k,\forall o \in V\}.
\end{equation} 
If $d_g^+=d_g^-$, which usually happens on physically interesting structures, we call them the growth exponent $d_g$, or the  {\it connectivity dimension}.

\noindent 
The connectivity dimension
$d_g$ is known for a large class of graphs: on lattices, it
coincides with the usual Euclidean dimension $d$, and for many fractals it has
been exactly evaluated \cite{gmh}. In general, we can think of it as the analogous of the fractal dimension, when the chemical distance metric is considered instead of the usual Euclidean metric.

\noindent
Infinite graphs are too general to describe systems of physical interest. 
Indeed, the discrete structures usually studied in physics are characterized by some
important properties, often implicitly assumed in literature, which can be translated in mathematical requirements: 
\begin{enumerate}
\item[\rm \bf A] We  consider only connected graphs,
since any physical model on
disconnected structures can be reduced to the separate study of the models
defined on each connected
component and hence to the case of connected graphs.
\item[\rm \bf B] Since physical interactions are always bounded,
the coordination numbers $z_i$, representing the number of neighbours interacting
with the site $i$, have to be bounded; i.e.
$\exists z_{max}\ |\ z_i\leq z_{max}\forall i\in V$.
\item[\rm \bf C] Real systems are always embedded in finite dimensional spaces.
This constraint requires for the graph $\cal{G}$ the conditions:
\begin{enumerate}
\item $\cal{G}$ has a polynomial growth (Definition \ref{growth})
\item
\begin{equation}
\lim_{r\to \infty} \frac{|\partial V_{o,r}|}{|V_{o,r}|}~=~0
\label{isop}
\end{equation}
where $\partial V_{o,r}$ denotes the border of $V_{o,r}$, i.e. the set of points of $V_{o,r}$ not belonging to $V_{o,r-1}$
(the existence itself of the limit is a physical requirement on $\cal{G}$). This condition is equivalent to require that boundary conditions are negligible in the thermodynamic limit.
\end{enumerate}
Notice that some graphs studied in physical literature, such as the Bethe lattice, do not satisfy (a) and (b), while many random graphs do not fulfil {\bf B}.
\end{enumerate}

\noindent
For a large class of physically interesting 
graphs we have considered so far, conditions (a) and (b) appear to be
equivalent. However for the equivalence of the two conditions a rigorous result
is still lacking.
A graph satisfying {\bf A}, {\bf B}  and
{\bf C} will be called {\it physical graph}. Conditions  {\bf A} and
{\bf B}  represent strong constraints on $\cal{G}$
and, as we will see later, they have very important consequences.

\section{Random walks on infinite graphs}

Considering random walks  on infinite structures, some further mathematical constraints are to be introduced to describe physical situations.

\noindent
First of all, the problem of uniform boundedness comes into play. Indeed, in (\ref{pijferr}) the conditions
\begin{equation}
\exists J_{min},J_{max}>0\ |J_{min}\leq\ J_{i,j}\leq J_{max} \quad \forall i,j
\label{limunif}
\end{equation}
together with {\bf B} are usually required to exclude the presence of a global bias, which would  generate a non-diffusive behaviour.

\noindent 
Moreover, in (\ref{wt}), in presence of waiting and traps, analogous considerations lead to the following conditions 
\begin{equation}
\exists w_{min},w_{max}>0\ |\quad{\rm either} \quad w_i= 0, \quad {\rm or} \quad  w_{min}\leq\ w_i\leq w_{max}\quad \forall i
\label{limunifw}
\end{equation}
\begin{equation}
\exists d_{min},d_{max}>0\ |\quad{\rm either} \quad d_i= 0, \quad {\rm or} \quad d_{min}\leq\ d_i\leq d_{max}\quad\forall i
\label{limunifd}
\end{equation}

\noindent
In the case of finite graphs, the possibility of associating to any matrix an operator acting on a finite dimensional vector space allowed to obtain very general and rigorous results.
In the infinite case, it is in general impossible to associate a linear operator acting on a Hilbert space to any matrix. However, when {\bf B} holds, the jumping probabilities matrix is quite particular: indeed, it only has a limited number of non vanishing entries in each row and column. Due to this property, the elements of a matrix product are given by finite sums, as in the finite graphs case, instead of being sums of series. Therefore, the typical convergence problems of infinite dimension space do not arise, allowing for a simple and effective study of random walks properties.

\noindent
Despite the increased mathematical complexity, many general results about infinite graphs have been rigorously proven. Some  have correspondents in the finite graphs case, but most of them concern quantities and properties which cannot be even defined on finite structures. The rest of this section is devoted to the former: following the same format used in section 5 we resume the main differences with respect to the finite case. The new properties arising in the thermodynamic limit will be discussed in the following sections.

\noindent
First of all, let us consider random walks  without traps on infinite graphs satisfying {\bf A},{\bf B}, (\ref{limunif}) and (\ref{limunifw}). 
It can be shown that 
\begin{equation}
P_{ij}^\infty=\lim_{t\to\infty}P_{ij}(t)=0\quad\quad\quad\quad \forall\quad i,j
\label{p limite 0}
\end{equation}
even for bipartite graphs. However, for bipartite graphs without staying probabilities, we still have
\begin{equation}
P_{ij}(t)=0 
\end{equation}
for all $t$ such that $t-r_{ij}$ is odd. Therefore, to study the large times asymptotic behaviours, as in the finite case we usually consider, for any given couple of sites $i$ and $j$, only the values $t_{ij}'$ of $t$ having the same parity as $r_{ij}$.
Unlike the finite case, the limit in (\ref{p limite 0}) in general is not reached exponentially. Indeed, if 
{\bf C} also holds, i.e. for physical graphs, the asymptotic behaviour is typically a power law, whose exponent only depends on topology, as we will discuss in details in the next sections. Notice that the widely studied case of Bethe lattices, not satisfying {\bf C}, is still characterized by an exponential decay.

\noindent 
Similarly, one can easily prove that 
\begin{equation}
\lim_{t\to\infty}F_{ij}(t)=0 \quad\quad\quad\quad\forall\quad i,j
\end{equation}
and
\begin{equation}
\lim_{t\to\infty}S_{i}(t)=\infty \quad\quad\quad\quad\forall\quad i
\end{equation}
the asymptotic behaviour being always bounded from above by $t$.

\noindent
As for the quantities concerning the number of visits and the first visit probabilities, the situation is far more complex. Indeed, dramatically different behaviours can occur, according to the graph topology.
In particular 
\begin{equation}
P_{ij}=\infty\quad\quad\quad\quad\forall\quad i,j \quad {\rm or}\quad <\infty \quad \forall\quad i,j
\label{pinf}
\end{equation}
\begin{equation}
F_{ij}=1\quad\quad\quad\quad\forall\quad i,j\quad {\rm or}\quad <1 \quad \forall\quad i,j
\label{finf}
\end{equation}
and
\begin{equation}
t_{ij}<\infty \quad\quad\quad\quad\forall\quad i,j \quad {\rm or}\quad =\infty \quad \forall\quad i,j
\end{equation}.

\noindent
The classification of infinite graphs according to these possible behaviours is the subject of the next section.

\section{Recurrence and transience: the type problem}

On finite graphs, in absence of traps, the probability of ever reaching (or returning to) a site, $F_{ij}$, is always 1. This means that the walker surely visits each site. This probability can be lowered only by adding traps, but in this case the total probability is not conserved, i.e. the walker asymptotically disappears. 
On infinite graphs, a third possibility arises, which is expressed in (\ref{pinf}) and (\ref{finf}): the walker can escape forever from its starting point, or never reach a given site, even in absence of traps. 

\noindent
This phenomenon was first noticed by Polya in 1921 on lattices: he showed that, while in $1$ and $2$ dimensions $F_{ij}=1$, for $d\ge 3$ $F_{ij}<1$ \cite{polya}. Since him, the former case has been called {\it recurrent} and the latter {\it transient}. Transience is an exclusive property of infinite graphs and it is fundamentally due to large scale topology. In other words, in the  transient case, it happens that the number of paths leading the walker away from its starting point is large enough, with respect to the number of returning paths, to act as an asymptotic trap (still conserving the total probability).

\noindent
As we will see in a while, transience and recurrence of random walks, when (\ref{limunif}) and (\ref{limunifw}) are satisfied, only depend on the graph topology. Therefore they are intrinsic properties of a discrete structure and the classification of infinite graphs according to them is also known as the  {\it type problem}.

\noindent
Let us define the problem mathematically.
First of all, a very general theorem on Markov chains states the following:
\begin{equation}
\exists\, i,j\in {\cal G} \quad| \quad F_{ij}=1\quad\quad \Rightarrow  \quad\quad F_{hk}=1\quad\forall\, h,k\in \cal{G}
\label{rictot}
\end{equation}
this means that recurrence is point independent, or, in other words, that if a walker surely reaches a point $j$ starting from a given point $i$, then it surely reaches any point $k$ starting from any point $h$. It is straightforward to see that an analogous result follows for the case $F_{ij}<1$. Therefore, recurrence and transience are global properties of a random walk.

\noindent
Another important result relates $F_{ij}$ and $P_{ij}$. Indeed, from (\ref{PP}), (\ref{FF}), and (\ref{gf2}), it follows that
\begin{equation}
\quad F_{ij}=1\quad\quad \Leftrightarrow \quad\quad P_{ij}=\infty
\label{PFric}
\end{equation}
and
\begin{equation}
\quad F_{ij}<1\quad\quad \Leftrightarrow \quad\quad P_{ij}<\infty
\label{PFtrans}
\end{equation}
i.e., a walk is recurrent (transient) if and only if any site is visited an infinite (finite) number of times. The latter can be taken as an alternative definition of recurrence and transience. However, as we will see, the situation is more complex when considering averages over all the sites.
A consequent property concerns the way the walker explores the sites of $\cal G$. Indeed, it can be shown that
\begin{equation}
\quad F_{ij}=1\quad\quad \Leftrightarrow \quad\quad \lim_{t\to\infty}{S_i(t)\over t} = 0
\label{Sric}
\end{equation}
while
\begin{equation}
\quad F_{ij}<1\quad\quad \Leftrightarrow \quad\quad 0< \lim_{t\to\infty}{S_i(t)\over t} <1
\label{Strans}
\end{equation}
In the first situation, where the number of distinct visited sites increases slower than the number of steps, is sometimes called {\it compact exploration}, since the subgraphs of the visited sites presents a negligible number of "holes".

\noindent
Recurrent graphs exhibit a further relevant property: one can show that

\begin{equation}
\lim_{\lambda\to 1^-}\,{\tilde{{P}_{ij}}(\lambda)\over \tilde{{P}_{hk}}(\lambda)}=\lim_{t\to\infty}\,{{{P}_{ij}}(t)\over {{P}_{hk}}(t)}= { z_j \over z_k } \qquad \forall\, i,j,h,k
\end{equation}

\noindent
The most important properties in the type problem concern its invariance with respect to a wide class of dynamical and topological transformations, establishing its independence of the graph details.

\noindent
First of all, consider two different random walks (without traps) on the same graph $\cal G$, one (W) defined by the ferromagnetic coupling matrix  $J_{ij}$  and by the waiting probabilities $w_i$,  and one (W') by $J'_{ij}$  and $w'_i$   .
It can be shown \cite{woesslibro} that, if both satisfy
 (\ref{limunif}) and (\ref{limunifw}), then W is recurrent if and only if W' is. 
In other words, any local bounded rescaling of ferromagnetic couplings and waiting probabilities leaves the random walk type unchanged. Therefore, provided the previously mentioned boundedness conditions are satisfied, the walk type only depends on the graph topology.

\noindent
Moreover, even the local topological details are irrelevant to determine the type of a graph. Indeed, it is possible to show that recurrence and transience are left invariant by adding and cutting of links satisfying the {\it  quasi-isometry} conditions.
More precisely, two graphs $\cal G$ and $\cal G'$ are called {\it quasi-isometric\/} if there are a mapping $\varphi: {\cal G} \to {\cal G'}$ 
and constants $A>0$, $B \ge 0$ such that
$$
A^{-1} r_{ij} - B \le r'_{\varphi i, \varphi j} \le A\,r_{ij} + B
$$
for all $i, j \in {\cal G}$, and
$$
r'_{ i', \varphi{\cal G} } \le B
$$
for every $i' \in {\cal G'}$. 

\noindent
If $B = 0$ then we say that $\cal G$ and 
$\cal G'$ are {\it metrically equivalent.\/} 
Quasi-isometries can be defined between arbitrary metric spaces and represent the most general local topology deformations.  Typical examples of them are given by the decimation transformations used on fractals and in real-space renormalization. In some sense, we can consider quasi-isometries as their extension to general networks.

\noindent
All the results presented so far refer to random walks without traps, i.e. to jumping probabilities given by (\ref{wt}) with $d_i=0$.
The introduction of at least one trap, setting  $d_i>0$ for at least one site $i$, has a very general and simple influence on the random walk behaviour: indeed transience is left unchanged, while recurrent random walks always become transient.

\section{The local spectral dimension}

As well as it happens for recurrence and transience properties, large scale topology affects the long time dependence of random walks quantities on infinite graphs. Indeed, it has been known for many years that, on regular (translation invariant) lattices, the exponents of the asymptotic power laws of random walks only depend on the lattice (Euclidean) dimension $d$. For example, 
\begin{equation}
 P_{ii}(t)\sim t^{-d/2}\qquad {\rm for}\quad {t\to\infty},\quad \forall i
\end{equation}
and
\begin{equation}
 S_{i}(t)\sim t^{min(1,d/2)}\qquad {\rm for}\quad {t\to\infty},\quad \forall i,\quad {\rm for}\quad d\neq 2
\end{equation}
(while, for $d=2$, $S_{i}(t)\sim t/\ln t )$. 
As we mentioned before, these laws typically present power behaviour even on general physical graphs, and the exponents of such powers can be used to define a generalized dimension.

\noindent
Let us consider a random walk without waitings and traps satisfying (\ref{limunif}), and suppose that, for a given $i\in \cal G$
\begin{equation}
 P_{ii}(t)\sim t^{-\widetilde d/2}\qquad {\rm for}\quad {t\to\infty}
\label{dtl}
\end{equation}
then it can be shown that 
\begin{equation}
 P_{hk}(t)\sim t^{-\widetilde d/2}\qquad {\rm for}\quad {t\to\infty},\quad \forall h,k
\end{equation}
(for bipartite graphs, the usual assumptions on the parity $t$ are understood).

\noindent
This means that the exponent of the power law is site independent and, therefore, it is a parameter characterizing the whole random walk. Since $\widetilde d = d$ on regular lattice, we can consider it as a dimension associated to the random walk on $\cal G$.
More precisely, we shall call {\it local spectral dimension} the limit
\begin{equation}
\widetilde d = -2\lim_{t\to\infty}{\ln P_{ii}(t)\over \ln t}
\label{defdtl}
\end{equation}
when it exists.

\noindent
Notice that the existence of this limit for a given $i$ implies it exists and has the same value for any $j\in \cal G$. Moreover the definition given in (\ref{defdtl}) is more general than (\ref{dtl}), since it includes the case of possible  multiplicative correction to the asymptotic behaviour, provided they are slower than any power law (e.g. logarithmic corrections).

\noindent
>From an historical point of view, the term "spectral dimension" was first introduced by Alexander and Orbach in 1982 \cite{aeo}, studying the anomalous vibrational dynamics on fractals. In the same work, they suggested that even the random walks should be ruled by the same parameter and wrote eq. (\ref{dtl}). Then the definition was generalized to general networks by Hattori, Hattori and Watanabe \cite{hhw}. Later, it has been shown that the anomalous dimension involved in vibrational dynamics is the {\it average spectral dimension} (\cite{debole}), we shall discuss in further sections, which coincides with $\widetilde d$ only for particular graphs, such as exactly decimable fractals. 

\noindent
As for the existence of the limit (\ref{defdtl}), a general theorem is still lacking, but it can be easily proven that the asymptotic decay of $ P_{ii}(t)$ is always bounded from above and from below by power laws. In any case, on all known cases of random walks on physical graphs, the local spectral dimension has been shown to exist. Notice that for the Bethe lattice, which does not fulfil the polynomial growth condition, the limit (\ref{defdtl}) is infinite.
>From now on, we shall consider random walks on graphs where $\widetilde d$ is defined.
Then, one can easily derive the following results:
\begin{itemize}
\item Random walks are recurrent if $\widetilde d<2$ and transient if $\widetilde d>2$. For $\widetilde d=2$, if (\ref{dtl}) holds, random walks are recurrent. However, subleading corrections to the power law can change the type to transient.
\item When (\ref{dtl}) holds,
\begin{equation}
  S_{i}(t)\sim t^{min(1, \widetilde d/2)}\qquad {\rm for}\quad {t\to\infty},\quad \forall i,\quad {\rm for}\quad \widetilde d\neq 2
\end{equation}
otherwise, in general,
\begin{equation}
  \lim_{t\to\infty}{\ln S_{i}(t)\over \ln t} = {min(1, \widetilde d/2)}\qquad \forall i,\quad {\rm for}\quad \widetilde d\neq 2
\end{equation}
\item When (\ref{dtl}) holds,
\begin{equation}
 F_{ij}(t)\sim t^{min(\widetilde d/2-2, -\widetilde d/2)}\qquad {\rm for}\quad {t\to\infty},\quad \forall i,j\quad {\rm for}\quad \widetilde d\neq 2
\end{equation}
otherwise, in general,
\begin{equation}
  \lim_{t\to\infty}{\ln F_{ij}(t)\over \ln t} = min(\widetilde d/2-2, -\widetilde d/2)\qquad \forall i,j\quad {\rm for}\quad \widetilde d\neq 2
\end{equation}
\end{itemize}

\noindent
The case $\widetilde d=2$ is rather particular. Indeed $\widetilde d=2$ is a critical dimension for random walks, discriminating recurrence from transience. The asymptotic behaviours of $S_{i}(t)$ and $F_{ij}(t)$ have a different dependence on $\widetilde d$ for $\widetilde d<2$ and $\widetilde d>2$. In particular, the probability of first visit has the same time decay of $P_{ij}$ for $\widetilde d>2$ while it decays faster for $\widetilde d<2$. When $\widetilde d=2$, the behaviours of $S_{i}(t)$ and $F_{ij}(t)$ are strongly affected by subleading corrections.

\noindent
As for the type problem, local spectral dimension presents interesting invariance properties.
First of all, it can be shown that waitings satisfying (\ref{limunifw}) do not affect its value \cite{debole}.
Moreover, for $\widetilde d<2$, on a given $\cal G$ it is the same for all ferromagnetic couplings satisfying (\ref{limunif}) \cite{hhw}. Unfortunately, an analogous result has not been proven for $\widetilde d>2$. However, we will see in later sections that the {\it average spectral dimension} has also this universality property.  

\noindent
The introduction of a finite number of traps do not affect $\widetilde d$ if $\widetilde d>2$. If $\widetilde d<2$ a finite number of traps (even only one) changes $\widetilde d$ to $\widetilde d + 1$. If the traps are infinite the behaviour is more complex and depends on their distribution.

\section{Averages on infinite graphs}

Usually, infinite graphs describing real systems are inhomogeneous, i.e., in mathematical terms, they are not invariant with respect to a transitive symmetry group. In simpler words, this means that the topology is seen in a different way from every site. The main effect of inhomogeneity is that the numerical values of physical quantities are site dependent. Therefore, one is typically interested in taking averages over all sites. This requires the introduction of suitable mathematical tools.

\noindent
First of all, the average in the thermodynamic limit $\bar{\phi}$ of a 
function $\phi_i$ defined on each site $i$ of the infinite graph $\cal{G}$ is defined by:
\begin{equation}
\overline{\phi}\equiv 
\lim_{r\rightarrow\infty} 
{\displaystyle \sum_{i\in S_{o,r}} \phi_i \over \displaystyle N_{o,r}}~~.
\label{deftd}
\end{equation}
The measure $|S|$ of a subset $S$ of 
$V$ is the average value $\overline{\chi(S)}$ of its characteristic 
function $\chi_i(S)$ defined by
$\chi_i(S)=1$ if $i\in S$ and $\chi_i(S)=0$ if $i\not\in S$.
The measure of a subset of links $E'\subseteq E$ is given by:
\begin{equation}
|E'| \equiv 
\lim_{r\rightarrow\infty} 
{\displaystyle E'_r \over  N_{o,r}}~~.
\label{mislinks}
\end{equation}
where $E'_r$ is the number of links of $E'$ contained in the sphere $S_{o,r}$.
The normalized trace $\overline{\rm Tr}B$ of a 
matrix $B_{ij}$ is:
\begin{equation}
\overline{\rm Tr}B \equiv {\overline b}
\end{equation} 
where $b_i \equiv B_{ii}$.
If condition {\bf C} holds, then
we can prove \cite{rimtim} that 
the averages of a bounded from below function $\phi_i$ are independent from 
the centre $o$ of the spheres sequence, using the fact that 
$\chi_i(S)$ is bounded and that measures of subsets are always well defined. 

\noindent
Now, due to this site independence, we have a good definition of averages which we will use in dealing with properties of random walks on infinite graphs.
As we shall see in the next section, on inhomogeneous networks the averages of site dependent functions can have a very different behaviour from their local counterparts, giving rise to rather unexpected phenomena.

\section{The type problem on the average}

In the last few years it has become clear that bulk properties are affected 
by the average values 
of random walks return probabilities over all starting sites: this is the 
case for spontaneous breaking of continuous symmetries \cite{mwg}, 
critical exponents of the spherical model \cite{cf}, harmonic vibrational 
spectra \cite{debole}. 
Therefore   
the classification of discrete structure in terms of {\it recurrence on 
the average} and {\it transience on the average} appears to be the most 
suitable. 
Unfortunately, while for regular lattices the two classifications are 
equivalent,
on more general networks they can be different and one has to study a 
Type-Problem
on the Average \cite{rimtim}. 

\noindent
This is defined using the return probabilities on the average 
$\bar{P}$ and $\bar{F}$, which are given by:
\begin{equation}
\bar{P}=\lim_{\lambda \rightarrow 1^-}\overline{\tilde{P}(\lambda)}\equiv\lim_{\lambda \rightarrow 1^-}\overline{\rm Tr}{\tilde{P}(\lambda)}
\label{dbP}
\end{equation}
\begin{equation}
\bar{F}=\lim_{\lambda \rightarrow 1^-}\overline{\tilde{F}(\lambda)}\equiv\lim_{\lambda \rightarrow 1^-}\overline{\rm Tr}{\tilde{F}(\lambda)}
\label{dbF}
\end{equation}
A graph $\cal{G}$ is called {\it recurrent on the average} (ROA) 
if $\bar{F}=1$, while it is {\it transient on the average} (TOA) 
when  $\bar{F}<1$.

\noindent
Recurrence and transience on the average are in general independent of the
corresponding local properties. The first example of this phenomenon 
occurring on inhomogeneous structures was found in a class of infinite
trees called NTD (see sect. \ref{mixed}) which are locally transient but recurrent on the 
average \cite{ntd}.

\noindent
Moreover, while for local probabilities (\ref{gf2}) gives:
\begin{equation}
\tilde{P}_{ii}(\lambda)\tilde{F}_{ii}(\lambda)\tilde{P}_{ii}(\lambda)+~1
\label{gf22}
\end{equation}
an analogous relation for (\ref{dbF}) and (\ref{dbP}) does not hold since 
averaging (\ref{gf22}) over all sites $i$ would involves the average of
a product, which due to correlations is in general different from the
product of the average. Therefore the double implication 
$\tilde{F}_i(1)=1 \Leftrightarrow \lim_{\lambda \rightarrow 1} 
\tilde{P}_i(\lambda)=\infty$ is not true. Indeed
there are graphs for which $\bar{F}<1$ but $\bar{P}=\infty$ (an example is
shown in Fig.1) and the 
study of the relation between $\bar{P}$ and $\bar{F}$ is a non trivial 
problem. 

\noindent
A detailed study of this relation  \cite{rimtim}
 shows that a complete picture of the behaviour of random walks 
on graphs can be given by dividing transient on the average graphs into 
two further classes, which are called {\it pure} and {\it mixed} 
transient on the average (TOA).

\noindent
First, considering a ROA graph, it can be proven that if $\bar{F}=1$ then 
$\bar{P}=\infty$. 
The proof can be easily generalized to graphs in which there is 
a positive measure subset $S$ such that:
$\lim_{\lambda\rightarrow 1}\overline{\chi(S)\tilde{F}(\lambda)}=|S|$.
Indeed in an analogous way it can be proven that:
\begin{equation}
\bar{P} \geq \lim_{\lambda\rightarrow 1}\overline{\chi(S')\tilde{P}(\lambda)}
=\infty~~~~ \forall S' \subseteq S, |S'|>0 
\label{PFg1}
\end{equation}

\noindent
We  call {\it mixed} transient on the average a TOA graph having a 
positive measure subset $S$ such that: 
\begin{equation}
\lim_{\lambda\rightarrow 1}\overline{\chi(S)\tilde{F}(\lambda)}=|S|.
\label{dmixed}
\end{equation}
while a graph is called {\it pure} TOA, if: 
\begin{equation}
\lim_{\lambda\rightarrow 1}\overline{\chi(S)\tilde{F}(\lambda)} < |S|~~~~
\forall S\subseteq V, |S|>0
\label{dpure}
\end{equation}

\noindent
Examples of pure TOA graphs are all the $d-$dimensional cubic lattices with $d>2$, while the  "haired cube" of Fig.1 is a typical mixed TOA graph.
Notice that a relevant theorem \cite{rimtim} establishes that for mixed TOA 
graphs we have $\bar{P}=\infty$, while for pure TOA graphs $\bar{P}$ is finite.
A further important property, characterizing 
mixed TOA graphs,  allows simplifying the study of statistical
models on these very inhomogeneous structures. 
It can be shown \cite{rimtim} that, in this case, the graph $\cal{G}$ can be always decomposed 
in a pure TOA subgraph $\cal{S}$ and a ROA subgraph $\bar{\cal{S}}$ with
independent jumping probabilities by 
cutting a zero measure set of links $\partial {\cal{S}} \equiv \{ (i,j)\in E |
i \in {\cal{S}} \and j \in \bar{\cal{S}} \}$.
The separability property implies that the two subgraphs
are statistically independent and that their thermodynamic  properties
can be studied separately. Indeed, in the thermodynamic limit, the partition functions referring 
to the two subgraphs factorize \cite{bcv}.
  
\noindent
To conclude this section, we note that the same invariance properties of the local type problem under addition of waiting probabilities, coupling rescaling and quasi-isometries still hold for the type problem on the average. This means, as for the local case, that recurrence and transience on the average are intrinsic properties of a graph and not only of a specific random walk defined on it.
On the other hand, the introduction of a finite number of traps does not change the type on the average.
Notice also that a slightly different definition of the type problem on the average can be found in mathematical literature \cite{milanesi}; it is more convenient for the formal development of the theory, but it is not directly related to statistical models on graphs.

\section{The average spectral dimension}

The asymptotic time dependence of the return probability on the average can be used to define a new intrinsic dimension which turns out to be very strictly related to the physical behaviour of statistical models on graphs \cite{mwg,debole,oenne}, as we will briefly discuss in the last section.

\noindent
Indeed, even if the asymptotic time decay of $P_{ii}(t)$ is always the same for all sites $i$, when the graph topology is strongly inhomogeneous it happens that its average over all the sites  decays according to a different law.
The average spectral dimension $\bar d$ is defined for physical graphs, like the local one in (\ref{dtl}) and (\ref{defdtl}), by
\begin{equation}
 \bar P(t)\sim t^{-\bar d/2}\qquad {\rm for}\quad {t\to\infty}
\label{dtm}
\end{equation}
when the asymptotic behaviour is a power law without subleading corrections, or, more generally, by
\begin{equation}
\bar d = -2\lim_{t\to\infty}{\ln \bar P(t)\over \ln t}.
\label{defdtm}
\end{equation}

\noindent
Notice, however, that, differently from the local case, no physical graphs are known, up to now, where the long time decay is not given by (\ref{dtm}).
Considerations analogous to those presented for the local case hold here, concerning the existence of the limit (\ref{defdtm}).
Obviously, in all cases where the local type is different from the average type, also the local spectral dimension differs from the average spectral dimension. A typical example, and, historically, the first one, is given again by NTD (see sect. \ref{mixed}) for a detailed account). 
However, the relations between $\bar d$ and the type problem on the average are not the same as in the local case. Indeed, while if $\bar d >2$ the walk is always pure TOA, random walks with $\bar d<2$ can be either pure ROA or mixed TOA.
 
\noindent
The most relevant property of $\bar d$ is without any doubt its strong invariance with respect to a very large class of dynamical and topological transformations, making it a unique universal parameter associated to a graph $\cal G$ \cite{debole,forte}.

\noindent
These transformations can be divided into three main classes: 
\begin{enumerate}
\item {\it Dynamical transformations leaving the graph topology unchanged.}
These consist in addition of waitings and of a finite number of traps, as well as in bounded local rescaling of ferromagnetic couplings.
\item {\it Topological transformations modifying the number of links but leaving the sites unchanged.}
These include "addition transformations" and "cutting transformations". The additions transforms consist in adding links joining sites up to an arbitrary but finite chemical distance from any site, while the cutting transform are defined to be their inverse.
The most general transformations consist in a combination of addition and cutting. Notice that even an infinite number of links can be 
modified with respect to the original graph.
\item {\it Topological rescaling, i.e. topological transformations modifying both links and sites.}
The most general topological rescaling
can be realized through two independent steps. The first one is the {\it
partition} and  consists in dividing 
the graph $\cal G$ in an infinite family of connected subgraphs $\cal G_\alpha$, with
uniformly bounded number of points.  
The second one is the {\it substitution} and consists in generating a new graph
$\cal G'$ by replacing some
or all $\cal G_\alpha$ by a different (connected) graph $\cal S_\alpha$, whose 
number of  points ranges from 1 to a fixed $N_{max}$, and by adding links 
connecting different $\cal S_\alpha$ in such a way that two generic $\cal S_\alpha$ 
and $\cal S_\beta$ are connected by some links if and only if $\cal G_\alpha$
and $\cal G_\beta$ were. 
The simplest topological rescaling occurs when every $\cal S_\alpha$ is composed by
just one point. In this case the resulting graph $\cal G_m$ is called the 
{\it minimal structure} of the partition $\{\cal G_\alpha\}$.
\end{enumerate}

\noindent
These three very general classes of geometrical transformations (together with even more general ones violating conditions {\bf  B} and {\bf C} and therefore not discussed here \cite{forte}  can be applied in all possible sequences to a graph, leading
to an overall transformation on coupling strength, number of links
and degrees of freedom  which does not change its spectral dimension $\bar d$.
We will call such a transformation an ${\it isospectrality}$. 

\noindent
Notice that isospectralities include quasi-isometries as a particular case.
Indeed, isospectralities include most part of currently used transformations.

\noindent
As an example, the usual decimation procedure on fractals 
is a topological rescaling. In particular, 
for all exactly decimable fractals (such as e.g. Sierpinski gaskets and 
T-fractals, as discussed in the next sections), 
the minimal structure of the graph coincides with the 
graph itself. Again, an isospectrality relates the usual two dimensional
square lattice, the hexagonal lattice and the triangular lattice,
which therefore all have dimension $2$.  
In other words, isospectralities are the theoretical formalization of the
intuitive idea of invariance with respect to bounded scale perturbations and
disorder and the {\it isospectrality classes}, defined as the classes of
graphs related by such transformations, are the practical realization of the
apparently abstract concept of non integer dimension.

\noindent
Now, since most dynamical and thermodynamical properties of generic discrete
structures depend only on $\bar d$, isospectralities provide a very powerful
tool to reduce a very complicated geometrical structure to the simplest one
having the same $\bar d$. The latter  turns out to be much simpler to study and
still presents the same universal properties.

\noindent 
Moreover, not only an isospectrality can be used to reduce and simplify
structures and problems. It can also be applied, with the opposite aim,  to
build complicated structures with controlled dynamical and thermodynamical
properties, starting from simple deterministic geometrical models. This is the
point of view of {\it spectral dimension engineering}, providing a very
interesting field of possibilities to polymer physicists and material
scientists dealing with non-crystalline materials.  
In Fig.2 we give explicit examples of isospectral structures obtained applying isospectral
transformations (without long range couplings) to the T-fractal
and to the  square lattice. 

\noindent
On macroscopically inhomogeneous graphs, it can happen that the average value of $P_{ii}(t)$ on infinite subgraphs of $\cal G$ with positive measure decays with a power law different from (\ref{dtm}) \cite{classi}. In such cases, it is interesting to look for the maximal (positive-measure) subgraphs having no (positive measure) parts with different power law decay. These are called {\it spectral classes} and each is characterized by its own spectral dimension. A theorem rather relevant in physical applications establishes that spectral classes can be separated from each other by cutting a zero-measure set of links, implying the same statistical independence property we discussed for mixed TOA graphs \cite{classi}.  

\section{A survey of analytical results on specific networks}
\label{survey}
Apart from the well known case of regular lattices, where it is completely
solved \cite{montroll}, the random walk problem    
has been studied analytically only on some specific classes of infinite graphs. In these cases,
one usually focuses on the asymptotic properties of random walks 
autocorrelation functions and on the calculation of the local and average spectral dimension. 
As we discussed in the previous sections, these are the most important quantities 
in statistical physics and thermodynamics. 
On lattices, the random walk problem is solved by using the translation invariance of the structure, and
this allows applying powerful mathematical tools, such as
the Fourier transform. On general graphs these methods do not apply. 
Therefore due to the lack of translation invariance, one has to introduce new 
and alternative techniques, which can be
grouped in three main classes: renormalization techniques, combinatorial techniques 
and mixed techniques. In the next subsections we will review recent and significant results obtained with
these techniques.

\subsection{Renormalization techniques}

Renormalization techniques have been successfully applied on deterministic fractals networks, where one can take advantage of the decimation transformations which connect two consecutive generations.  In particular, a well studied class of fractals is that of {\it exactly decimable} fractals. On these structures, exact renormalization group 
calculations based on a real space decimation procedure allow obtaining all the relevant random walks quantities.

\noindent
Let us consider a random walk without
traps and sources defined by the jumping probabilities (\ref{pij}) and let us write the master equation for the probability 
$P_{0i}$ of being at site $i$ after $t$ steps for a random walker starting from an 
origin site $0$ at time $0$:
\begin{equation}
P_{0i}(t+1) - P_{0i}(t)= \sum_j  A_{0j}({P_{0j}(t)\over z_j} - {P_{0j}(t)\over z_i}) + \delta_{i0}\delta_{t0}
\label{3}
\end{equation}
Equation (\ref{3}) can be written in terms
of the generating function $\widetilde P_{ij}(\lambda)$ when $\lambda \to 1^-$
by setting $\lambda = 1-\epsilon $, writing:
\begin{equation}
\widetilde P_{ij}(\epsilon) = \sum_{t=0}^{\infty}(1+\epsilon)^{-t} P_{ij}(t)
\label{4}
\end{equation}
and taking $\epsilon =\to 0$:
\begin{equation}
\epsilon\widetilde P_{0i}(\epsilon) =\sum_j A_{0j}
({\widetilde P_{0j}(\epsilon)\over z_j} - {\widetilde P_{0j}(\epsilon)\over z_i})+\delta_{0i}.
\label{5}
\end{equation}
Notice that the system (\ref{5}) is inhomogeneous and corresponds
to a Cauchy problem, which has only one solution. The behaviour of such  
solution 
for  $\epsilon\to 0$ is what we need to obtain the local spectral dimension $\widetilde d$,
as defined in (\ref{dtl}), through the Tauberian Theorems. On the other hand, to calculate the average
spectral dimension $\bar d$ we will need to average 
over all starting points the solution of equation (\ref{5}), strongly modifying its asymptotic behaviour on inhomogeneous graphs, as we will see in the following. 

\noindent
Exactly decimable fractals are a restricted class of self similar structures
(i.e. not all self similar structures are exactly decimable)
which are geometrically invariant under site decimation. This invariance 
is explicitly applied in analytical calculations for random walks.
A geometrical structure is decimation invariant if it is possible to
eliminate a subset of points (and all the bonds connecting these points) 
obtaining a network with the same geometry of the starting one. 
>From a mathematical point of view this 
corresponds to the possibility of eliminating by substitution a set of equations
from system (\ref{3}) or (\ref{5}) obtaining a system which is similar to the 
initial one after a suitable redefinition of the coupling parameters.
Examples of exactly decimable fractals are the Sierpinski Gasket (Fig.3), 
\cite{ret,ret1,ret2,ret3}, the $T-$fractal, shown in Fig.4 \cite{redner,ses}, 
the branched Koch curves, in Fig.5 \cite{koch}.
In general, all deterministic finitely-ramified fractals  are
exactly decimable. Notice that exact decimation is a particular case of isospectrality,
as we discussed in previous sections. 

\noindent
Let us consider now the general procedure to decimate the set of equation (\ref{3}).
After eliminating a set of points and substituting the corresponding equation,
one finds:
\begin{equation}
\epsilon \rightarrow \epsilon'(\epsilon)\sim a^2\epsilon
\label{8}
\end{equation}
The presence of the term $\delta_{i0}$ in (\ref{5}) requires a 
redefinition of the quantities $\widetilde P_{ij} (\epsilon)$ to assure that, 
even after the decimation, the initial condition will correspond to the 
probability of being in a fixed site equal to 1.
One introduces a new parameter $c$ and writes the transformation 
law for $\widetilde P_{ij} (\epsilon)$ as:
\begin{equation}
\widetilde P_{ij} (\epsilon) \rightarrow \widetilde P'_{ij} (\epsilon')
\sim {1\over c} \widetilde P_{ij} (\epsilon)
\label{9}
\end{equation}
From the rescaling of $\epsilon$ and $\widetilde P_{ij} (\epsilon)$, the local spectral dimension  $\widetilde d$ is obtained by using a suitable expression for $\widetilde P_{00} (\epsilon)$:
\begin{equation}
\widetilde P_{00} (\epsilon) \sim \epsilon ^{ \widetilde d /2 -1}
\label{10}
\end{equation}
which holds only for $\widetilde d<2$. This is always the case for exactly decimable fractals.
Using expression (\ref{10}) one easily finds:
\begin{equation}
\widetilde d = 2 \, {{ \log a^2/c}\over {\log a^2}}
\label{12}
\end{equation}

\noindent
As for the average spectral dimension $\bar d$ , by using the relation between equation 
(\ref{5}) and the equation for harmonic oscillations to be discussed later \cite{split}, one has that
\begin{equation}
\bar d = { \log r \over \log a}
\label{13}
\end{equation}
where $r$ is the decimation ratio used in the renormalization procedure.
Therefore $\widetilde d =\bar d$ if 
\begin{equation}
r=a^2/c
\label{18}
\end{equation}
This can be shown to be the case for all exactly decimable fractals, using
results obtained \cite{hhw} for the Gaussian model.

\noindent
Equation (\ref{13})
allows calculating the spectral dimension on all exactly
decimable fractals, once the decimation procedure is identified, recovering
known results.

\noindent
One of the most studied fractal is without any doubt the Sierpinski gasket
\cite{ret,ret1,ret2,ret3} and its generalizations. For the simplest
case one has $r=3$ and $a=\sqrt 5$, leading to:
\begin{equation}
\widetilde d = 2 \,{{ \log 3}\over {\log 5}}
\label{dsierp}
\end{equation}
For d-dimensional generalized Sierpinski gaskets, which are built from
a $d-$ dimensional hypertetrahedron of side length $b$ filled with $b$ layers
of smaller hypertetrahedra of unit site length, Hilfer and Blumen \cite{ret1} have shown that
for $b=2$
\begin{equation}
\widetilde d = 2 \,{{ \log d+1}\over {\log d+3}}
\label{blumen}
\end{equation}
and for $b=3$
\begin{equation}
\widetilde d = 2 \,{{ \log ((d+1)(d+2)/2)}\over {\log ((d+2)(2d^2+9d+19)/(4d+6))}}
\label{blumen2}
\end{equation}
 
\noindent
Due to the self-similarity of the structure, the return probabilities on
the Sierpinski gasket show a remarkable effect, which has been pointed out in
\cite{woessierp}. Indeed, the coefficients have an oscillatory behaviour, which is given
by:
\begin{equation}
P_{00}(t) = t^{-\widetilde d /2} F({\log t \over \log 5})
\label{sierposc}
\end{equation}
where $F$ is a periodic $C^\infty$-function of period 1 whose Fourier series is given by
\begin{equation}
F(x) = \sum_{k=-\infty}^\infty \Gamma (1- {log 3\over log 5} +  {2 \pi k i\over log 5})^{-1}
exp(2 \pi k i x)
\label{sierposc1}
\end{equation}
\noindent
Interestingly, it can be shown that the oscillation of the coefficients disappears in the probability of 
return on the average.

\noindent
The renormalization techniques can be applied to all exactly decimable fractals.
For example, for the $T-$fractal \cite{redner,ses}, which is a particular case of hierarchical combs \cite{giac}, 
one has $r=3$ and $a=\sqrt 6$.

\subsection{Combinatorial techniques}

Renormalization procedures cannot be applied on non self-similar graphs. Therefore one has to develop
alternative techniques to study the random walk problem.
This is the case of bundled structures \cite{fibrati1,fibrati2}, a large class of very interesting graphs   
used in condensed matter as realistic models for the geometry and dynamics of polymers and other inhomogeneous
systems. 
Given two graphs $\cal B$ and $\cal F$ , not necessarily different, and a site $F$ of $\cal F$ , we call bundled graph with base $\cal B$ and fibre $\cal F$ the graph  built by joining to each site of $\cal B$ a copy of $\cal F$ in such a way that $F$ is the only site $\cal B$ and $\cal F$  have in common (Fig.6). Examples
of bundled structures are comb polymers \cite{degennes} (Fig.7), brush polymers, shown in  Fig.8, and many kinds of branched aggregates (Fig.9). For these graphs a purely combinatorial technique allows calculating of the
asymptotic properties of the random walk autocorrelation functions.

\noindent
Let us consider a walker starting from a point belonging to the base and let us
restrict ourselves to base graphs with constant coordination number $z_{\cal B}$. By decomposing
the motion of the walker on the fibre and on the base, one can obtain: 
\begin{equation}
P_0(t)=\sum_{t_{_{\cal B}}=0} ^\infty \sum_{t_1=0} ^\infty ...\sum_
{t_{_{\cal B}}+1=0} ^\infty P_{_{\cal B}}(t_{_{\cal B}})\left({z_{_{\cal B}}
\over {z_{_{\cal B}}+z_{_{F}}}}\right)^
{t_{_{\cal B}}} P_{_{\cal F}}'(t_1)
\cdots P_{_{\cal F}}'(t_{t_{_{\cal B}}+1})\delta_{t,t_{_{\cal
B}}+\sum_{i=1}^{t_{_{\cal B}}+1} t_i}
\label{darling}
\end{equation}
where $P_{_{\cal F}}'$ refers to a random walk on $\cal F$ with a trap in the
starting point of $\cal F$. In terms of the generating functions equation, (\ref{darling}) becomes:
\begin{equation}
 \widetilde{P}_0(\lambda)=\sum_{t_{_{\cal B}}=0}^{\infty}P_{_{\cal B}}(t_{_
{\cal B}})\left({\lambda z_{_{\cal B}} \over {z_{_{\cal B}}+z_{_{F}}}}
\right)^{t_{_{\cal B}}}\left( \widetilde{P}_{_{\cal F}}'(\lambda)\right)^
{t_{_{\cal B}}+1}
=\widetilde{P}_{_{\cal F}}'(\lambda) \widetilde{P}_{_{\cal B}}(\lambda')
\label{pretty}
\end{equation}
with
\begin{equation}
\lambda'\equiv {\lambda z_{_{\cal B}}\over {z_{_{\cal B}} +z_{_{ F}}}}
\widetilde{P}_{_{\cal F}}'(\lambda)
\end{equation}
and
\begin{equation}
\widetilde{P}_{_{\cal F}}'(\lambda)=\left( 1-{z_{_{F}}\over {z_{_{\cal B}}+
z_{_{F}}}}
\left(1-\left(\widetilde{P}_{_{\cal F}}(\lambda)\right)^{-1}\right) 
\right)^{-1}
\label{heaven}
\end{equation}
with $\widetilde{P} _{_{\cal F}} (\lambda)$ being the generating function
of the probability of returning to the starting point $F$ on $\cal F$ 
without the trap. From this relations one obtains the values for the
local spectral dimension on general bundled graphs:
\begin{equation}
\widetilde{d}=\cases{ \widetilde{d}_{_{\cal F}}& if $\widetilde{d}_{_{\cal 
F}}
\ge  2$ \cr
\vspace{.05 cm} \cr
4-\widetilde{d}_{_{\cal F}}   & if $\widetilde{d}_{_{\cal F}} \le  2$ e
$\widetilde{d}_{_{\cal B}} \ge 4$
 \cr
\widetilde{d}_{_{\cal F}} + \widetilde{d}_{_{\cal B}} -
{\widetilde{d}_{_{\cal F}} \widetilde{d}_{_{\cal B}}
\over 2} & if $\widetilde{d}_{_{\cal F}} \le  2$
 e $\widetilde{d}_{_{\cal B}}\le 4$
\cr}
\label{miro}
\end{equation}
where $\widetilde {d}_{_{\cal B}}$ and $\widetilde {d}_{_{\cal 
F}}$ are the local spectral dimension of the base and of the fibre.
If the coordination number of the base is not constant, it can be shown that
this amounts to introduce waiting probabilities on the points connecting the
fibre and the base, which, as shown in the previous section, does not change
the value of the spectral dimension.

\noindent
As for the average spectral dimension, it is easy to show that if the fibre is 
an infinite graph, the average spectral dimension of the whole graph is the 
spectral dimension of the fibre. On the other hand, if the fibre is a finite graph,
the average spectral dimension coincides with that of the base.

\noindent
>From equations (\ref{pretty}), (\ref{heaven}) one also obtains the asymptotic laws
for the probability of returning to the starting point, which on these structures 
can contain logarithmic corrections. 
Indeed, writing
\begin{equation}
 P_0(t) \sim  \prod_{i=0} ^{\infty} {}^i\ln^{\beta(i)} (t)
\end{equation}
and setting
\begin{equation}
m= {\rm min}
\{i\ge 0 | \beta(i) \neq -1\}
\end{equation}
and
\begin{equation}
I\left( \widetilde {d} / 2 \right) = \cases { 1 & if 
$\widetilde {d} / 2 $ is an integer \cr
 0 & otherwise \cr }
\end{equation}
one has:

a) if $\widetilde {d}_{{\cal B}} <4$ and $\widetilde {d}_{{\cal 
F}} <2$
\begin{equation}
 \beta(i)=\cases{ -1 & for $0<i<m$  \cr
\vspace{.2 cm}\cr 
\left( 1\!-\!{{\widetilde{d}_{_{\cal B}}} \over 2} \right)\left[
\beta_{_{\cal F}}(m_{_{\cal F}}) + I\left({{\widetilde{d}_{_{\cal F}}} 
\over 2}
\right)
\right]\!-\!I\left({\widetilde{d}_{_{\cal F}}\over 2}\right) &  for $i=mm_{_{\cal F}}$  \cr
\vspace{.3 cm} \cr
\left(1\!-\!{{\widetilde{d}}_{_{\cal B}} \over 2 }\right)
\beta_{_{\cal F}}(i) +
\theta \big(i\!-\!m_{_{\cal F}}\!-\!m_{_{\cal B}}\!\big) \beta_{_{\cal B}}
\big(i\!-\!m_{_{\cal F}} \big)+
&   \cr
\vspace{.1 cm} \cr
 \hfill + \delta_{_{i-m_{_{\cal F}},m_{_{\cal B}}}}I\left(
{\widetilde{d} \over 2}\right)\! -\! \delta_{_{i,m}}I\left(
{\widetilde{d}\over 2} \right)                   &  otherwise  \cr }
\end{equation}
where $m_{{\cal B}}$ and $m_{{\cal B}}$ refers to the base and to the fibre
respectively while $m$ refers to the whole graph and is determined by: 
\begin{equation}
m =m_{_{\cal F}} + \delta_{\widetilde{d}_{_{\cal B}},2}
m_{_{\cal B}}
\end{equation}

b) if $\widetilde {d}_{{\cal B}} >4$ and $\widetilde {d}_{{\cal 
F}} <2$
\begin{equation} 
\beta(i)=\cases{  - \beta_{_{\cal F}}(i) -2\delta_{i,m_{_{\cal F}}}
I({\widetilde{d}_{_{\cal F}} /2 }) & for $i\ge m_{_{\cal F}}$ \cr
\beta_{_{\cal F}}(i) & for $0<i<m_{_{\cal F}}$  \cr }
\end{equation}

c) if $\widetilde {d}_{{\cal B}} =4$ and $\widetilde {d}_{{\cal 
F}} <2$ and $d_{{\cal B}}> -1$ \\

$\beta(i)$ has to be determined as in a).\\

d) if $\widetilde {d}_{{\cal B}} =4$ and $\widetilde {d}_{{\cal 
F}} <2$ e $m_{{\cal B}}<-1$ \\

$\beta(i)$ has to be determined as in b).\\

e) if $\widetilde {d}_{{\cal F}} >2$ 
\begin{equation}
\beta(i)=\beta_{_{\cal F}}(i) \,\,\,\,\,\,\, \forall i
\end{equation}
The case $\widetilde {d}_{{\cal F}} =2$ has to be treated separately, 
as the case $\widetilde {d}_{{\cal F}} <2$ if the fibre is a recurrent graph 
or as the case $\widetilde {d}_{{\cal F}} 
>2$ if it is transient.

\noindent
Another interesting way of combining together two graphs to obtain a more complex
structure is the {\it Cartesian product}. The Cartesian product of 
two graphs $X, Y$ has vertex set $X \times Y$, and two pairs $xy$, 
$x^{\prime}y^{\prime}$ are adjacent if $x \sim x^{\prime}$ and
$y=y^{\prime}$,
or $x=x^{\prime}$ and $y \sim y^{\prime}$. An example of an interesting Cartesian product
is that of the Toblerone graph \cite{toble}, shown in Fig.10, which is obtained from the product of 
a line with a Sierpinski gasket. Using combinatorial techniques analogous to those presented
for bundled graphs, it can be show that the local and the average spectral dimension on
the whole graph are the sum of the corresponding dimensions of the two initial graphs \cite{woesslibro}.

\subsection{Mixed techniques}
\label{mixed}
The random walk problem on some very interesting cases of graphs cannot be studied simply by one of
the above cited techniques and it requires instead a "mixed" use of the two, which gives rise
to very interesting phenomena. Indeed, the first example of a difference between the
local and the average spectral dimension, the "dynamical dimension
splitting",  was observed on the quasi self-similar graphs $NT_D$, where the asymptotic properties 
of the random walk were found by a mixed technique \cite{ntd}.
 
\noindent
The fractal trees known as $NT_D$ \cite{doyle} can be recursively defined as
follows:  an origin point $O$ (Fig.11)
is connected to a point $1$ by a link, of unitary 
length; from $1$, the tree splits in $k$ branches of length $2$ (i.e. 
consisting of two consecutive links); the ends of these
branches split in $k$ branches of length $4$ and so on; each
endpoint of a branch of length $2^n$ splits in $k$ branches of length $2^{n+1}$.

\noindent
As one can easily verify, $NT_D$ are not exactly decimable  and therefore the
simple decimation techniques cite above cannot be applied. Indeed, after a
simple decimation starting from the origin $O$, one obtains $k$ copies of the
original structure joined together in a point instead of the same $NT_D$.
However, $NT_D$ are invariant under a more complex transformation $T=D\cdot C$, consisting of the product of a cutting transform $C$ and a decimation $D$, that can be
described as follows.
Let us cut the log of the tree in point $1$ and separate the $k$ 
branches (cutting transform). Now, each branch can be obtained
from the initial $NT_D$ by a dilatation with a factor 2.
Eliminating all branches but one and decimating it 
(decimation transform), one obtains the original $NT_D$.

\noindent 
The $T$ transform can now be used to solve the random 
walks problem. Let us sketch the main points of the calculation.
The cutting transform gives a  relation between random
walks on the whole tree and random walks on one of its branches; more precisely 
one relates $\widetilde P_{O} ^{tree} (\lambda)$, the generating function of 
the probability of returning to point $O$ after a random walk on the $NT_D$ 
tree, and $\widetilde P_{1} ^{branch} (\lambda)$, the generating function of
the probability of returning to the starting point $1$ after a random walk on
one  of the branches. This relation is given by \cite{ntd}:
\begin{equation}
\widetilde P_{O} ^{tree}(\lambda) ={ {\widetilde P_{1} ^{branch} (\lambda)
+k} \over {2\lambda \widetilde P_{1} ^{branch} (\lambda) +k}}
\label{24}
\end{equation}
Now, the decimation transformation is performed using a 
time-rescaling technique. Indeed, the motion of the 
random walker on the branch considered only after an even number of steps
can be exactly mapped in the motion of a random walker on the tree after 
the introduction of a staying probability $p_{ii}=1/2$ in every site $i$. 
This equivalence can be translated in terms of generating functions through the 
substitutions:
 \begin{equation}
  \widetilde{P_O}(\lambda)\rightarrow {\lambda \over {2 -\lambda}}
\widetilde{P_O}\left(  {2 \over {2 -\lambda}} \right)
\label{uno}
\end{equation}
\begin{equation}
 \lambda \rightarrow  \lambda^2
\label{due}
\end{equation}
Equations (\ref{uno}) and (\ref{due}) can be used to rewrite 
(\ref{24}) as:
\begin{equation}
\widetilde{P_O}^{tree}(\lambda)={{
{2 \over {2 -\lambda ^2}}\widetilde{P_O}^{tree}\left(
{\lambda ^2 \over { 2- \lambda ^2}}\right) +k} \over {
( 1- \lambda^2) {2 \over {2 -\lambda ^2}} \widetilde{P_O}^{tree}\left(
{\lambda ^2 \over { 2- \lambda ^2}}\right) +k}}
\end{equation}
Choosing a suitable power law expression for the singularity of $P_{OO}^{tree}(\lambda)$ for $\lambda \to 1^-$ 
\cite{ntd}
we obtain:
\begin{equation}
\widetilde d = 1+ {{\log k} \over {\log 2}}
\label{27}
\end{equation}

\noindent
To obtain the average spectral dimension, one has to calculate the normalized trace
of the return probability $\widetilde{P_O}^{tree}(\lambda)$. It can be shown 
that ${\overline d}=1$ and this can be 
intuitively understood noting that the topology of $NT_D$ 
is dominated by linear chains which become longer and longer 
in the outer branches \cite{split}. Therefore, while $NT_D$ are locally transient if the ramification $k$ is
greater that $2$, they are always recurrent on the average. This result has been generalized.
Indeed, recently it has been shown that all physical trees, satisfying conditions {\bf A}, {\bf B} and {\bf C}
are recurrent on the average \cite{donetti}.

\noindent
The cutting decimation transform can be applied to a large class of non-exactly decimable fractals
which correspond to more general cases of the $NT_D$. This are built with the same recurrence procedure
as the $NT_D$ and we shall call them
$2^mNT_D$, $nNT_D$ and $p-polygon NT_D$, depending on the growth rules for the branches \cite{ntdgen}.

\noindent
The first generalization is that of $2^mNT_D$.
The $2^mNT_D$ are infinite fractal trees that can be recursively built using 
the same recipe as for $NT_d$ but, from point $1$, the log splits in $k$ branches of 
length $2^m$ (i.e. made of $2^m$ consecutive links)
which, in turn, split in $k$ branches of length $2^{2m}$  
and so on in such a way that each branch of length $2^{nm}$ 
splits in $k$ branches of length $2^{(n+1)m}$. The case $m=1$ corresponds to
the usual $NT_D$ previously studied. 
If $m>1$ the time rescaling procedure which led to (\ref{uno}) and (\ref{due}) must be iterated 
$m$ times obtaining:
\begin{equation}
\widetilde{P_O}^{tree}(\lambda)={{
\left(
\prod_{i=1} ^{m} {2 \over {2 -\lambda_i ^2}} \right)
\widetilde{P_O}^{tree}(\lambda
_{i+1})+k} \over {(1-\lambda^2)\left( \prod_{i=1} ^{m} {2 \over {2 -\lambda_i 
^2}}  \right)\widetilde{P_O}^{tree}(\lambda _{i+1})+k}}
\label{muir}
\end{equation}
with
\[
\lambda_i = \cases  {\lambda& $i=1$ \cr
                      {} \cr
                     {{\lambda_{i-1}^2}\over {2- \lambda_{i-1}^2}}
  & $i>1$  \cr}
\]
$i$ being the iteration step.
This gives, with the same steps as for $m=1$: 
\begin{equation}
\widetilde{d}_{2^m}=1 + {{\ln k} \over{\ln 2^m}}
\end{equation}
which represent the generalization of the result obtained for $m=1$.

\noindent
The previous results can be extended to $nNT_D$,
where now $n$ is an integer and not necessarily a power of 2, and to 
$p\!-\!polygon NT_D$, where the branches of $NT_D$ 
are replaced by $p$-vertices regular polygons (Fig.12). 

\noindent
Let us consider $nNT_D$ first.
While relation (\ref{24}) for the cutting transform still holds, the exact time-rescaling 
procedure can not be applied to the branch of generic length $n$.
However even in this case it is possible to obtain an asymptotic recursion
relation applying the Renormalization Group techniques usually implemented 
on exactly decimable fractals. Although this procedure cannot give
an exact equation for $\widetilde{P_O}^{tree}(\lambda)$ as in the previous case,
nevertheless it can be used to obtain the exact value of $\widetilde{d}$
via an asymptotic estimation.

\noindent
Indeed, in this case the branch of the $nNT_D$ can be considered as
a tree with a dilatation factor equal to $n$. 
The log of this tree can be reduced to a unitary length log after the 
suppression of the $n-2$ sites between the edges and introducing a new link 
connecting the edges. The same operation can be repeated for branches of every 
length suppressing the inner $n-2$ consecutive sites in every sequence of $n$ 
sites and introducing a new link between the surviving points. The final 
structure is equal to the original tree and the generating function  
$\widetilde{P_1}^{branch}(\lambda)$ becomes 
$\widetilde{P_1}^{'branch}(\lambda')$ where:
\begin{equation}
\lambda' = n^2 \lambda
\label{tre}
\end{equation}
\begin{equation}
\widetilde{P_1}^{'branch}(\lambda')= { 1\over n} 
\widetilde{P_1}^{branch}(\lambda)
\label{quattro}
\end{equation}
Now $\widetilde{P_1}^{'branch}(\lambda')$ coincides with
$\widetilde{P_O}^{tree}(\lambda')$ since our branch has been transformed 
into a tree and (\ref{24}) can be rewritten as:
\begin{equation}
\widetilde{P_O}^{tree}(\lambda) ={{ n \widetilde{P_O}^{tree}(n^2 \lambda) +k}
\over{2\lambda
 n \widetilde{P_O}^{tree}(n^2 \lambda) +k}}
\label{stella}
\end{equation}
Using the procedure described in the previous section for $2^m NT_D$, 
from (\ref{stella}) it follows that for an $n-NT_D$ the spectral dimension 
is given by:
\begin{equation}
\widetilde{d}_{n}=1 + {{\ln k} \over{\ln n}}
\end{equation}
An analogous technique can be used for $p\!-\!polygon NT_D$ (Fig.12).
The log polygon has now $p$ faces of unitary length; from each of $p-1$ of its 
vertices $k$ polygons depart, whose faces have length $n$ and so on. These 
structures, though similar to $NT_D$ are no longer loopless structures nor 
necessarily bipartite graphs (e.g. the $3\!-\!polygon$ tree).
The Cutting-Decimation transform 
can be applied to $p\!-\!polygon NT_D$ as 
in the case of $NT_D$ with the same substitutions (\ref{tre}) and (\ref
{quattro}). Indeed, even if (\ref{24}) does not hold
in this case, a new relation 
between the generating functions of the tree and that of one of its branches
can be obtained using bundled structures theory discussed above \cite{dieci}. 
Let us consider a 
$p\!-\!polygon NT_D$ and suppose to attach 
$k$ branches also in the free vertex of the log
(the root of the tree): 
we obtain a bundled structure having the log polygon as base and 
the graph made of $k$ branches as fibre. 
Since for a $p$-polygon:
\begin{equation}
\widetilde{P_O}(\lambda) \sim {1 \over{ p (1- \lambda)}}
\end{equation}
as $\lambda \rightarrow 1$, we obtain for our bundled structure:
\begin{equation}
\widetilde{P_O}^{b.s.}(\lambda)={1 \over{1-{k\over{k+1}}\widetilde{F_1}
^{branch}(\lambda)}}{1\over p} \left(
1- { \lambda\over {k+1}}{1 \over { 1- 
{k\over{k+1}}\widetilde{F_1}^{branch}(\lambda)}} \right) ^{-1}
\label{primo}
\end{equation}
where $\widetilde{P_O}^{b.s.}(\lambda)$ is the generating function of the 
probability of returning to point $O$ (one of the vertices of the log polygon)
after a random walk on the bundled structure and $\widetilde{F_1}
^{branch}(\lambda)$ is the generating function of the probability of returning 
for the first time
to the point of connection with the base after a random walk on the fibre. 
Now,
\begin{equation}
\widetilde{F_O}^{b.s.}(\lambda)= {k \over {k+1}}
\widetilde{F_1}^{branch}(\lambda)+ {1\over {k+1}}\widetilde{F_O}^{tree}
(\lambda)
\label{terzo}
\end{equation}
where 
${F_O}^{tree}(\lambda)$ refers to the $p\!-\!polygon NT_D$. From (\ref{primo}), 
and (\ref{terzo}) and using the usual relation between $\widetilde{F_1}
^{branch}(\lambda)$ and $\widetilde{P_1}
^{branch}(\lambda)$, a relation between 
$\widetilde{P_O}^{tree}(\lambda)$ and $\widetilde{P_1}^{branch}(\lambda)$
follows, which represents the cutting transformation.  
It is now possible to perform the Cutting-Decimation transform
to $p\!-\!polygon NT_D$ and get:
\begin{equation} 
\widetilde{d}_{p}=1 + {{\ln k(p-1)} \over{\ln n}}
\end{equation}
In the same way we can calculate the spectral dimension of an $NT_D$ built
with $d-$dimensional simplexes instead of $p\!-\!polygons$. A $d$-dimensional
simplex is a complete graph of $d+1$ points i.e. a graph where each point
is nearest neighbour of all other points. The $2-$dimensional case is the
triangle, the $3-$dimensional one is the tetrahedron and so on.
Since for $d-$simplex $\widetilde{P_O}(\lambda)\sim 1/(d+1)(1-\lambda)$
the spectral dimension is:
\begin{equation}
\widetilde{d}_{d}=1 + {{\ln kd } \over{\ln n}}
\end{equation}

\section{Relation with other physical problems}

As we have shown in previous chapters, the random walk problem is strictly related to graph topology.
Indeed, the main physical quantities are simple functions of the adjacency matrix $A$, which algebraically describes the graph structure. Now, the Hamiltonians of a series of fundamental statistical models are linear in $A$, therefore even their behaviour is deeply influenced by topology and it can be expressed in terms of random walks functions. Due to this reason, the main concepts and parameters characterizing random walks, such as recurrence and transience, as well as the spectral dimension, determine also the properties of these models, which have very different physical origins. This provides a very powerful tool to investigate and classify geometrically disordered and inhomogeneous systems, where the usual techniques and ideas developed for lattices do not apply.  

\subsection{The oscillating network}

Probably, the physical model whose connection with random walks has been most extensively explored is the so-called {\it oscillating network}.

\noindent
The harmonic oscillations of a generic network  
of masses $m$ linked by springs of elastic constant $K$ can be studied 
by writing the equations of motion of the displacements $x_i$ of each mass
from its 
equilibrium position:
\begin{equation}
m {d^2 \over dt} x_i = - K \sum_{j} A_{ij}(x_i-x_j)= - K \sum_{j} \Delta_{ij}x_j
\label{equosc}
\end{equation}
which after Fourier transforming with respect to the time reads:
\begin{equation}
{\omega^2 \over \omega_0^2} \tilde{x}_i = \sum_{j} \Delta_{ij}\tilde{x}_j  
\label{equoscft}
\end{equation}
where $\omega_0^2\equiv K/m$. In other words, the determination of the normal modes and of the normal frequencies of the oscillating network reduces to the diagonalization of the Laplacian operator $\delta$.

\noindent
Noticing that $\Delta=Z( {\bf 1}-P)$, where {\bf 1} the identity matrix and $P$ is given by (\ref{pij}), it is not difficult to establish mathematical correspondences with random walks.
In particular, using the universality properties discussed in the previous sections, one can show a fundamental result concerning the density $\rho(\omega)$ of normal modes at low frequencies:
\begin{equation}
\rho(\omega) \sim \omega^{\bar d-1}\qquad {\rm for}\quad {\omega\to 0}
\label{rodiomega}
\end{equation}

\noindent
This basic connection between random walks and harmonic oscillations was first introduced by Alexander and Orbach in 1982 for the case of fractals. Notice that at that time the splitting between local and average spectral dimension on inhomogeneous structures was not yet known and  the exponent describing the scaling of the density of states at low frequencies was simply called spectral dimension, since it was related to the vibrational spectrum.
Due to the already mentioned universality properties, the above result hold for the very general case where oscillating masses and elastic constants may have different values on different sites and links, provided they are bounded by positive numbers. More precisely, considering the equations of motions
\begin{equation}
m_i {d^2 \over dt} x_i = - K\sum_{j} J_{ij}(x_i-x_j)= - K \sum_{j} L_{ij}x_j
\label{equoscgen}
\end{equation}
for the same  graph of (\ref{equosc}), if (\ref{limunif}) holds together with
\begin{equation}
\exists m_{min},m_{max}>0\ |m_{min}\leq\ m_{i}\leq m_{max} \quad \forall i
\label{limunifm}
\end{equation}
then the asymptotic behaviour of the density of vibrational states is still given by 
(\ref{limunif}).

\noindent 
>From all the above properties, it follows that also the spectrum of 
the Laplacian operator $L$ depends on $\bar d$: indeed it can be shown \cite{debole} that 
the spectral density $\rho(l)$ of $L$ at low
eigenvalues behaves as  $\rho(l)\sim l^{~{\bar d}/2 -1}$.

\noindent
The average spectral dimension is crucial in determining the behaviour of the oscillating network in equilibrium with a thermal bath at temperature $T$. Considering the Hamiltonian of the system given by (\ref{rodiomega})
\begin{equation}
 H=\sum_{i}{p_i^2\over 2 m_i} + {1\over 2 } m \omega_0^2 \sum_{ij} \,J_{ij}x_i x_j
\label{hamosc}
\end{equation}
and calculating the thermodynamic averages with respect to the Gibbs weight $\exp(-H/kT)$, where $k$ denotes the Boltzmann constant,
one can show that, for positive $T$, 
\begin{equation}
\bar{<x^2>}=\infty\qquad {\rm for}\quad \bar d \le 2
\label{pldiv}
\end{equation}
while
\begin{equation}
\bar{<x^2>}<\infty\qquad {\rm for}\quad \bar d > 2.
\label{plconv}
\end{equation}
This is the generalization to graphs of the fundamental Peierls result about the thermodynamic instability of oscillating crystals in low dimensions. In other words, for an infinite oscillating network with $\bar d \le 2$ in equilibrium with a thermal bath, the mean square displacement of masses from their equilibrium positions would diverge. 

\subsection{The Gaussian model}

The Gaussian model is the simplest statistical model used to study magnetic systems on lattices. Even if it is not realistic, its properties are fundamental to understand more complex and phenomenologically significant models. In field theory it is also known as "free scalar field". 
The Gaussian model on $\cal G$ is defined by the Hamiltonian:
\begin{equation}
H = {1\over 2}\sum_{ij} \phi_i(J L_{ij}+m_i^2\delta_{ij})\phi_j -h\sum_i \phi_i
\label{ham}
\end{equation}
where $\phi_i$ is a real field, $J>0$ a ferromagnetic coupling, $h$ an external
magnetic field and $m_i^2=\alpha_i m^2$, with $1/K<\alpha_i<K$ for some
positive $K$\cite{debole}. Its specific free energy $f_G$ is given 
by  
\begin{equation}
f_G(J,m_i^2,h) = \lim_{N\to \infty} {1\over N} ~F = -\lim_{N\to \infty} 
{1\over N} \log Z
\end{equation}
where $Z$ is the partition function calculated according to the
Boltzmann weight $\exp(-H)$. 
The spectral dimension is related to the singular part of $f_G$ for
$h=0$ and $m^2\to 0$ by:
\begin{equation}
Sing~(f) \sim m^{\bar d}.
\label{sing}
\end{equation}   The covariance of this Gaussian process reads
\begin{equation}
\langle \phi_i \phi_j \rangle \equiv  C_{ij}(m^2) = 
(\Delta + m^2 \eta)^{-1}_{ij}
\label{gauss2}
\end{equation}
and hence it satisfies by definition the Schwinger--Dyson (SD) equation 
\begin{equation}
\label{eq:SD}
	(J_i + m^2\eta_i) C_{ij}(m^2) - \sum_{k\in G} J_{ik} C_{kj}(m^2) 
	= \delta_{ij}
\end{equation} 
Setting
\begin{equation}
	C_{ij} = {(1-W)^{-1}_{ij} \over {J_i + m^2 \eta_i}} ,~~~
	W_{ij} = {J_{ij}\over {J_j + m^2 \eta_j}} 
\end{equation}
one obtains the standard connection with the random walk (RW) over
$\cal G$ \cite{hhw}:
\begin{equation}
\label{eq:RW}
	(1-W)^{-1}_{ij} \,=\,\sum_{t=0}^\infty \, (W^t)_{ij} = 
	\sum_{\gamma:\,i\leftarrow j} W[\gamma]
\end{equation}
where the last sum runs over all paths from $j$ to $i$, each weighted
by the product along the path of the one--step probabilities in $W$:
\begin{equation}
	\gamma = (i,k_{t-1},\ldots,k_2,k_1,j) \Longrightarrow
	W[\gamma] = W_{ik_{t-1}} W_{k_{t-1}k_{t-2}},\ldots,W_{k_2k_1}W_{k_1j}
\end{equation}
Notice that, as long as $m>0$, we have $\sum_i (W^t)_{ij}<1$ for any
$t$, namely the walker has a non-zero death probability. This implies
that $C_{ij}$ is a smooth functions of $m^2$ for $m\ge\epsilon>0$. In
the limit $m\to 0$ the walker never dies and the sum over paths in
eq. (\ref{eq:RW}) is dominated by the infinitely long paths which sample
the large scale structure of the entire graph (``large scale'' refers
here to the metric induced by the chemical distance alone). This
typically reflects itself into a singularity of $C_{ij}$ at $m=0$ whose
nature does not depend on the detailed form of $J_{ij}$ or $\eta_i$,
as long these stay uniformly positive and bounded.

\noindent
Of particular importance is the leading singular infrared behaviour, as $m^2\to
0$, of the average $[C(m^2)]_G$ of $C_{ii}(m^2)$, which is a positive
definite quantity,  over all points
$i$ of the graph ${\cal G}$, which we may write in general as
\begin{equation}
	{\rm Sing}\, [C(m^2)]_G  \sim c (m^2)^{{\bar d}/2-1} 
\label{leadsing}
\end{equation}

\subsection{Spherical model and $O(n)$ models}

The spherical model is again a magnetic model with no direct connection to phenomenology. Nevertheless, is a little more complex than the Gaussian one and, most important, it exhibits phase transitions at finite temperature for $\bar d >2$. Moreover, its critical exponents can be exactly determined and they turn out to be simple functions of $\bar d$, pointing out the crucial role of the average spectral dimension in phase transitions and critical phenomena.
The spherical model can be defined on a generic graph through the Hamiltonian
(\ref{ham}) with the {\it generalized spherical constraint}
$
\sum_i z_i \phi_i^2 = N
$.
We assume the 
coordination numbers to be bound:
$1 \le z_i \le z_{\max}$. 
Its free energy and correlation functions can be expressed in terms of the Gaussian ones. Then the critical behaviour is obtained from the infrared singularities of the latter, i.e. in terms of the long time behaviour of random walks.
The results concerning the critical exponents are summarized in the following table, where $T_c=0$ for $\bar d\le 2$:
\newcommand{\displayfrac}[2]{\frac{\displaystyle #1}{\displaystyle #2}}
\begin{table}[htb]
{\centering
\begin{tabular}{|l|c|c|c|}
\hline
& $1 \le \bar{d}<2$
& $2<\bar{d}<4$
& $\bar{d}>4$\\
\hline
\hline
{\tiny \ } &  &  & \\
$T=T_c$
& $\delta \to \infty$
& $\delta= \displayfrac {\bar{d}+2} {\bar{d}-2}$
& $\delta=3$\\
{\tiny \ } &  &  & \\
\hline
\hline
{\tiny \ } &  &  & \\
$T<T_c$
& -
& $\gamma'$ does not exist
& $\gamma'=1$\\
{\tiny \ } &  &  & \\
\hline
{\tiny \ } &  &  & \\
$T>T_c$
& $\gamma= - \displayfrac {2} {\bar{d}-2}$
& $\gamma= \displayfrac {2} {\bar{d}-2}$
& $\gamma=1$\\
{\tiny \ } &  &  & \\
\hline
\hline
{\tiny \ } &  &  & \\
$T<T_c$
& -
& $c= \displayfrac {1}{2} K_B$
& $c= \displayfrac {1}{2} K_B$\\
{\tiny \ } &  &  & \\
\hline
{\tiny \ } &  &  & \\
$T>T_c$
& $\alpha= \displayfrac {\bar{d}} {\bar{d}-2}$ 
& $\alpha= \displayfrac {\bar{d}-4} {\bar{d}-2}$
& $\alpha=0$\\
{\tiny \ } &  &  & \\
\hline
\hline
{\tiny \ } &  &  & \\
$T<T_c$
& -
& $\beta=\displayfrac {1}{2}$
& $\beta=\displayfrac {1}{2}$ \\
{\tiny \ } &  &  & \\
\hline
\end{tabular}
\caption{Critical exponents of the spherical model on a graph of 
spectral dimension $\bar{d}$}
\par}
\end{table}

\noindent
The so called $O(n)$ models are defined, for positive integer $n$,  by the  
Boltzmann weight $\exp(- \beta H_{n})$, where
\begin{equation}
H_{n}[{\bf S}]  = {1\over 2} \sum_{<ij>} J_{ij} ({\bf S}_i - {\bf S}_j)^2
\label{oennem}
\end{equation}
the sum extends to all links of a certain graph ${\cal G}$,   $J_{ij} > 0$ are ferromagnetic interactions, which may vary
from link to link, and ${\bf S}_i$ is an $n-$dimensional vector of
fixed length normalized by ${\bf S}_i\cdot{\bf S}_i =n$. 
They represent more realistic magnetic models, but their exact solution is in general impossible.
However, a series of complex but powerful inequalities, relating their correlation functions to the random walks generating functions, allow proving some very general results shedding light on the complicated phenomena concerning phase transitions on graphs. 
In particular it has been proven that:
\begin{itemize}
\item they cannot have phase transitions at $T>0$ if ${\cal G}$ is recurrent on the average;
\item they exhibit phase transitions at $T>0$ if ${\cal G}$ is transient on the average;
\item for $n\to\infty$ their critical exponents tend to the spherical ones.
\end{itemize}

\end{document}